\documentclass[10pt,singlecolumn,
superscriptaddress,
english,
prx,
showpacs,
floatfix,
aps,
longbibliograph]{revtex4-2}

\usepackage{xcolor} 

\usepackage{graphicx}
\usepackage{ntheorem}
\usepackage{mathrsfs}
\usepackage{subfigure}
\usepackage{dcolumn} 

\usepackage{hyperref}
\usepackage{comment}
\usepackage{bbm}

\usepackage{amsmath}
\usepackage{amssymb}
\usepackage{longtable}
\usepackage{boldline}
\usepackage{braket}

\usepackage{siunitx}

\usepackage{placeins}

\usepackage{array}

\begin{document}
	\title{Quantum heat engine with long-range advantages}
	
	\author{Andrea Solfanelli}
	\email{asolfane@sissa.it}
	\affiliation{SISSA, via Bonomea 265, I-34136 Trieste, Italy}
	\affiliation{INFN, Sezione di Trieste, Via Valerio 2, 34127 Trieste, Italy}
	
	\author{Guido Giachetti}
	%\email{ggiachet@sissa.it}
	\affiliation{SISSA, via Bonomea 265, I-34136 Trieste, Italy}
	\affiliation{INFN, Sezione di Trieste, Via Valerio 2, 34127 Trieste, Italy}
	
	\author{Michele Campisi}
	%\email{michele.campisi@nano.cnr.it}
	\affiliation{NEST, Istituto Nanoscienze-CNR and Scuola Normale Superiore, I-56127 Pisa, Italy}
	%\address{Department of Physics and Astronomy, University of Florence, I-50019, Sesto Fiorentino (FI), Italy}
	%\affiliation{INFN - Sezione di Pisa, I-56127 Pisa, Italy}
	
	\author{Stefano Ruffo}
	%\email{ruffo@sissa.it}
	\affiliation{SISSA, via Bonomea 265, I-34136 Trieste, Italy}
	\affiliation{INFN, Sezione di Trieste, Via Valerio 2, 34127 Trieste, Italy}
	\affiliation{\mbox{
			Istituto dei Sistemi Complessi,
			Via Madonna del Piano 10, I-50019 Sesto Fiorentino, Italy
	}}
	
	\author{Nicol\`o Defenu}
	\email{ndefenu@phys.ethz.ch}
	\affiliation{Institut f\"ur Theoretische Physik, ETH Z\"urich, Wolfgang-Pauli-Str. 27 Z\"urich, Switzerland}

	\date{\today} 
	\begin{abstract}
		{The employment of long-range interactions in quantum devices provides a promising route towards enhancing their performance in quantum technology applications.
			Here, the presence of long-range interactions is shown to enhance the performances of a quantum heat engine featuring a many-body working substance. We focus on the paradigmatic example of a Kitaev chain undergoing a quantum Otto cycle and show that a substantial thermodynamic advantage may be achieved as the range of the interactions among its constituents increases.  
			Interestingly, such an advantage is most significant for the realistic situation of a finite time cycle: the presence of long-range interactions reduces the non-adiabatic energy losses, by suppressing the detrimental effects of dynamically generated excitations. This effect allows mitigating the trade-off between power and efficiency, paving the way for a wide range of experimental and technological applications.}
	\end{abstract}
	\maketitle 
	\section{Introduction}
	%
	%
	%Thermodynamics was born in the nineteenth century from practical considerations, namely, understanding the functioning of thermal machines such as heat engines and refrigerators. Since then, the performances of such devices have experienced outstanding 
	%improvements following the needs of modern technological applications that utilize thermal energy, ranging from fuel-based vehicles, household air conditioners, and refrigerators to the most advanced dilution refrigerators\,\cite{ZuCryogenics2022}.
	Thermodynamics was born during the industrial revolution in response to the very practical necessity  to understand and master the functioning of thermal machines such as heat engines and refrigerators. This has led to the development of thermal machines of increasingly high performance in terms of thermodynamic efficiency and power consumption. Such devices permeate our daily life and are as well crucial for technology advances. Few out of countless examples span from fuel-based vehicles, household air conditioners and refrigerators to the dilution refrigerators for quantum applications\,\cite{ZuCryogenics2022}.
	Fault-tolerant quantum computing\,\cite{Preskill18QUANTUM2} represents the latest technological challenge, which in recent years has attracted significant research efforts\,\cite{RaymerQST2019,RiedelQST2019,YamamotoQST2019,SussmanQST2019,RobersonQST2019,AielloQST2021,RobersonQST2021}. In particular the route towards fault tolerance of available quantum processors is dictated by the quantum threshold theorem\,\cite{AharonovProceedings97,KitaevRussianMathSurvey97,KnillScience98}, which states that it is possible to correct errors, even by using noisy gates, provided that the noise level remains below a certain threshold. 
	
	Cooling quantum hardwares at sufficiently low temperatures allows, in principle, to achieve this threshold, but the introduction of large classical apparata, i.e. thermal baths, may generate additional sources of decoherence. Thus, the design of microscopic and coherent thermodynamic machines constitutes an urgent technological challenge\,\cite{AuffevesPRXQuantum2022}, which has led researchers to study quantum thermal engines, i.e., heat engines and refrigerators whose working substance operates directly in the quantum domain\,\cite{MyersAVSQuantumScience2022}. Much theoretical\,\cite{AlickiJPA1979,AllahverdyanPRE2008,Campisi15NJP17,Campisi16NatCommun7,KarimiPRB2016,FriedenbergerEPL2017,KosloffEntropy2017,Deffner19Book} and experimental\,\cite{BrantutScience2013,RossnagelScience2016,VanHorneNPJ2020,KlaersPRX2017,VonLindenfelsPRL2019,PetersonPRL2019,KlatzowPRL2019,BoutonNature2021} efforts have been devoted to their study, recently showing their potential applicability to existing quantum processors\,\cite{SolfanelliPRXQuantum2021,SolfanelliAVSQuantum2022}. However these devices suffer from the well-known trade off between power and efficiency\,\cite{Deffner19Book}. This is due to two main contributions: the first one comes from the fundamental limitations imposed by the second law of thermodynamics on irreversible processes, which implies that the thermodynamic efficiency of a heat engine has to be smaller than that of a Carnot engine\,\cite{Fermi1956}. Furthermore, any realistic cycle is carried out in a finite time. This limitation is an additional source of losses, due to the dynamic generation of excitations in the system which dissipate energy, worsening the performance of the device. Consequently, an increase in power typically involves a greater dissipation, with detrimental effects on efficiency. Various methods that suppress the non-adiabatic transitions, based on the so-called shortcuts to adiabaticity, have been proposed to overcome this problem\,\cite{ChenPRL2010,CampoSciRep2014,AbahEPL2017,GueryRevModPhys2019}. However these techniques typically require switching on additional driving fields, a procedure whose additional energetic cost reduces the actual work output of the device\,\cite{CampbellPRL2017,AbahPRE2018}. 
	
	Here, we propose a new paradigm to reduce these detrimental effects, which consists in using, as the working substance of the engine, a long-range interacting quantum system, i.e., a system in which the coupling energy between two of its microscopic constituents $J_{i,j}$  decays as a power law of their distance $r = |i-j|$: $J_{i,j}\propto r^{-\alpha}$, with $\alpha>0$\,\cite{Campa2014,DefenuArXiv2021}.
	
	Long-range interacting systems are emerging as promising platforms for quantum technological applications, due to their stability against external perturbations, which allows keeping the impact of dynamically generated excitations under control, therefore mitigating their detrimental effects\,\cite{xuPhysics2022}. An example of the rigidity of long-range interacting platforms against external drivings, is the possibility for such systems to host clean discrete Floquet time crystal phases\,\cite{RussomannoPRB2017,SuracePRB2019,PizziNatComm2021,GiachettiArXiv2022}, i.e., robust out-of-equilibrium symmetry broken phases in systems subjected to an external time-periodic driving\,\cite{ElseAnnRevCondMattPhys2020}. Moreover, long-range interactions are known to alter the universal behaviour of critical systems at and out of equilibrium, generating a wide range of unprecedented phenomena. These include novel form
	of dynamical phase transitions\,\cite{DefenuPRB2019,HalimehPRR2020}, defect formation\,\cite{AcevedoPRL2014,HwangPRL2015,DefenuPRL2018,DefenuPRB2019}, anomalous thermalization\,\cite{RegemortelPRA2016}, information spreading\,\cite{TranPRX2020,ChenPRL2019,KuwaharaPRX2020} and metastable phases\,\cite{DefenuPNAS2021,GiachettiArXiv2021}, whose features have no-counterpart in systems with short-range interactions.

	In this work we show how the peculiar properties of long-range interacting quantum systems may be used to boost the performances of a many-body quantum Otto cycle\,\cite{KosloffEntropy2017,Deffner19Book}, which we take as a prototypical example of a quantum thermal device. In particular, we identify multiple advantages stemming from the presence of long-range interactions
	\begin{enumerate}
		%This phenomenon may be explained in terms of the universal dynamical scaling of the density of the excitations\,\cite{ZurekPRl2005,DamskiPRA2006,PolkovnikovRevModPhys2011,DelCampo2014,DziarmagaAdvancesinPhysics2010,DeGrandi2010}, which is affected by the presence of long-range interaction\,\cite{DefenuPRB2019}. Since such scaling does only depend on universal properties, we expect this result to hold regardless of the microscopic details of the model. 
		\item In the limit of an infinitely slow cycle, enhanced optimal performances of the device are observed, confirming previously reported results\,\cite{Wang20PRE102}. In particular, the enhancement is observed in the engine most useful operation modes: the heat-engine mode and the refrigerator mode.  
		\item  In the realistic case of a finite time cycle featuring the crossing of a quantum critical point, the choice of a long-range-interacting working substance leads to a significant reduction in nonadiabatic energy losses and defect proliferation if compared with its short-range counterpart.
	\end{enumerate}
	
	From the experimental point of view, long-range interacting systems may be implemented in typical quantum simulation platforms, such as atomic molecular and optical (AMO) systems\,\cite{BrittonNature2012,MonroeRevModPhys2021,MottlScience2012,MivehvarAdvancesinPhysics2021,ChomazArXiv2022}. %Interestingly trapped ions setups allow to tune the exponent $\alpha$ of the power law decay interaction between $\alpha\simeq 0-3$\,\cite{BrittonNature2012}.
	Interestingly, trapped ions setups allow to tune the power law exponent $\alpha$, dictating the decay of the interaction energy with distance, from $\alpha \simeq 0$ to $\alpha \simeq 3$\,\cite{BrittonNature2012}.
	Moreover, such platforms have been proven useful in realising quantum computers\,\cite{CiracPRL95,KielpinskiNature2002,SchindlerNJP2013,FriisPRX2018,PogorelovPRXQuantum2021,IonQ}. Finally, single atom quantum heat engines have been already implemented in trapped ions setups\,\cite{RossnagelScience2016}. Accordingly, the implementation of thermodynamic devices featuring long-range interacting working substances may be feasible in current trapped ion platforms, with possible direct applications to the cooling of the existing ion-based quantum processors\,\cite{IonQ}.
	
	The paper is organized as follows. In Sec.~\ref{sec: Many-body quantum Otto cycle} we introduce the many-body Otto cycle, taking as a prototypical example of a long-range working substance, the long-range version of the celebrated Kitaev chain\,\cite{KitaevUFN2001}. In Sec.~\ref{sec: adiabatic cycle} we focus on the limit of the infinite-time, adiabatic, Otto cycle, observing a long-range advantage in the optimal regime. Finally, in Sec.~\ref{sec: Finite time cycle} we estimate the energy losses of more realistic finite-time cycles with respect to the adiabatic case, showing that they are reduced in the critical regime thanks to the presence of long-range interactions. 
	
	%generic quadratic one dimensional fermionic model with, possibly, long-range interactions, also known as \,\cite{KitaevUFN2001,MaityJPA2019,VodolaPRL2014,VodolaNJP2015,LeporiAnnalsofPhysics2016,UhrichPRB2020,DefenuPNAS2021,DefenuArXiv2021}.
	
	%
	%
	\section{Many-body quantum Otto cycle}\label{sec: Many-body quantum Otto cycle}
	\subsection{Many body working substance}
	As a prototypical example of a many-body working substance with, possibly, long-range interactions, we consider a generic model of spinless fermions hopping across the $N$ sites of a linear chain in the presence of pairing interaction, and with a chemical potential $h$. Assuming periodic boundary conditions, the system Hamiltonian reads
	\begin{align}
		H = &-\sum_{j=1}^N\sum_{r=1}^{N/2-1}\left[t_r c_{j+r}^\dagger c_j+\Delta_rc_{j+r}^\dagger c_j^\dagger +h.c.\right]\notag\\
		&-h\sum_{j=1}^N\left[1-2c_j^\dagger c_j\right],\label{eq: long-range Kitaev chain H}
	\end{align}
	\begin{figure*}%[t!]
		\centering
		\includegraphics[width=\textwidth]{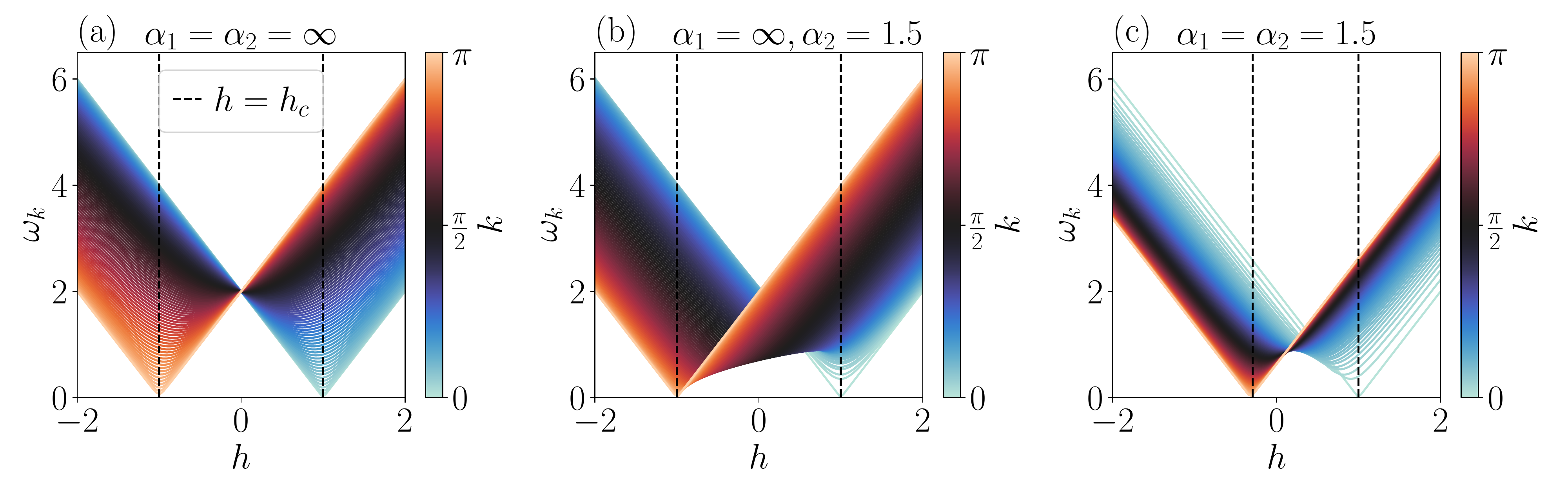}
		\caption{Single particle spectrum for different $k$-modes, corresponding to different colors, as a function of the chemical potential $h$. Different panels correspond to diffent values of $\alpha_1$,$\alpha_2$, in particular: a) nearest neighbor pairing and hopping with $\alpha_1 = \alpha_2 = \infty$; b) long-range pairing and nearest neighbor hopping with $\alpha_1 = \infty$, $\alpha_2 = 1.5$; c) long-range pairing and hopping with $\alpha_1 = \alpha_2 = 1.5$. Black dashed vertical lines indicate the quantum critical points $h = h_c$. The size of the system is $N = 200$.}
		\label{fig: spectrum}
	\end{figure*}
	where $c_j^\dagger$ and $c_j$ are creation and annihilation operators for fermions at site $j$, while $t_r$ and $\Delta_r$ are the hopping and pairing amplitudes, respectively. We choose their dependence on the intersite distance $r$ according to the power laws
	\begin{align}
		t_r = \frac{1}{N_{\alpha_1}}\frac{J}{r^{\alpha_1}},\quad \Delta_r = \frac{1}{N_{\alpha_2}}\frac{\Delta}{r^{\alpha_2}},
	\end{align}
	with the hopping exponent $\alpha_1>0$, the pairing exponent $\alpha_2>0$,  and $N_\alpha = \sum_{r=1}^{N/2}r^{-\alpha}$ the Kac scaling factor\,\cite{KacJMP1963}, which guarantees extensivity of the energy in the case $\alpha_i<1$, with $i = 1,2$. Hereafter, we set $J = \Delta = 1$ as the energy scale and work in units of $\hbar = k_B = 1$. This model, often referred to as long-range Kitaev chain\,\cite{KitaevUFN2001,VodolaPRL2014,MaityJPA2019}, has been mainly studied in two particular cases: with short-range hopping $\alpha_1 = \infty$ and generic long-range pairing $\alpha_2$\,\cite{VodolaPRL2014,VodolaNJP2015,LeporiAnnalsofPhysics2016}, and with same range for  pairing and hopping $\alpha_1 = \alpha_2 = \alpha$\,\cite{DefenuPNAS2021,DefenuArXiv2021}. As observed in Refs.\,\,\cite{jaschke2017critical,vanderstraeten2018quasiparticle,defenu2019dynamical}, the latter case can be related to the quantum Ising model. In particular, in the short-range case with $\alpha\to\infty$, the relation becomes exact through the Jordan-Wigner mapping\,\cite{Fradkin2013}. In this Section, we review the main properties of the model with general power law couplings as in Eq.~\eqref{eq: long-range Kitaev chain H}. %  then in the rest of the paper we will mainly focus on the second on the equally long-range case, since as shown in Appendix \ref{app: Characterization of the long-range advantage} this provides more significant long-range advantages in quantum thermodynamics applications thanks to the peculiar properties of its spectrum. 
	For reasons that will become clear later, we restrict ourselves to the regime $\alpha_1,\alpha_2>1$. 
	
	Due to the translational invariant nature of the couplings, it is useful to write the Hamiltonian in terms of the momentum-space operators 
	\begin{align}
		\tilde{c}_k = \frac{e^{-i\frac{\pi}{4}}}{\sqrt{N}}\sum_{j=1}^N e^{ikj}c_j,
	\end{align}
	where $k = 2\pi n/N$, with $n = -N/2+1,\dots, N/2$ (In the following we will drop the $\tilde{}$ on the $c_k$ unless it is ambiguous). Then we obtain 
	\begin{align}
		H = \sum_k\large[(h&-t_k)(c_k^\dagger c_k-c_{-k}c_{-k}^\dagger)\notag\\
		&+\Delta_k(c_k^\dagger c_{-k}^\dagger+c_{-k}c_{k})\large],
	\end{align}
	where $t_k$ and $\Delta_k$ are the Fourier transforms of the hopping and pairing amplitudes, respectively, which in the thermodynamic limit may be written as
	\begin{align}
		&t_k = \mathrm{Re}\left[\mathrm{Li}_{\alpha_1}(e^{ik})\right]/\zeta(\alpha_1),\label{eq: hopping}\\
		&\Delta_k = \mathrm{Im}\left[\mathrm{Li}_{\alpha_2}(e^{ik})\right]/\zeta(\alpha_2),\label{eq: pairing}
	\end{align}
	where $\mathrm{Li}_{\alpha}(z)$ denotes the polylogarithm and $\zeta(\alpha)$ is the Riemann zeta function. We notice that the Hamiltonian in the Fourier space can be decomposed into the sum of single mode Hamiltonians, introducing $\Psi_k = (c_k,c^\dagger_{-k})^T$
	%of $2\times2$ Hamiltonians $\mathcal{H}_k$, one for each $k$-mode, in a basis given by $\Psi_k = (c_k,c^\dagger_{-k})^T$
	\begin{align}
		&H =\sum_{k}\Psi_k^\dagger\mathcal{H}_k\Psi_k,\\ \label{eq:Hk}
		&\mathcal{H}_k = (h-t_k)\sigma_k^{z}+\Delta_k\sigma_k^x. 
	\end{align}
	where $\sigma_k^{(a)}$, $a=x,y,z$ are the sigma Pauli operators. Let us notice how the $k$-th term of the Hamiltonian acts on a different sector of the total Hilbert space, namely the two dimensional subspace spanned by the states $|0_{k},0_{-k}\rangle $, $|1_{k},1_{-k}\rangle = c_k^\dagger c^\dagger_{-k}|0_{k},0_{-k}\rangle$.
	Then the Hamiltonian is diagonalized via a Bogoliubov transformation, in terms of the fermionic quasiparticle operators $\gamma_k = u_kc_k+v^*_{-k}c_{-k}^\dagger$, with Bogoliubov coefficients 
	\begin{align}
		u_k = \cos\frac{\theta_k}{2},\quad v_k = \sin\frac{\theta_k}{2},\label{eq: bogoliubov coefficients}
	\end{align}
	where $\theta_k=\arctan[\Delta_k/(h-t_k)]$, to obtain
	\begin{align}
		H = \sum_k\omega_k(h)\left(\gamma_k^\dagger\gamma_k-1/2\right),
	\end{align} \label{eq:Hdiag}
	with the spectrum 
	\begin{align}
		\omega_k(h) = 2\sqrt{(h-t_k)^2+\Delta_k^2}
	\end{align}\label{eq: spectrum}
	For $\alpha_1,\alpha_2>1$, the system possesses two different phases separated by two quantum critical points $h_{c} =1,-1+2^{1-\alpha_1}$, in correspondence of which the dispersion relation becomes gapless near to the critical mode $k_c = 0,\pi$, respectively\,\cite{UhrichPRB2020,DefenuArXiv2021}. Figure \ref{fig: spectrum} shows the single-particle spectrum $\omega_k$ for different values of $0\leq k\leq \pi$, plotted as a function of the chemical potential $h$. In particular Fig. \ref{fig: spectrum}a) shows the spectrum for a short-range Kitaev chain, which, as previously stated, is exactly equivalent to the nearest neighbor quantum Ising chain, showing a ferromagnetic and an antiferromagnetic quantum critical point located at $h_c = 1,-1$, respectively. A similar structure is present in the two relevant long-range cases displayed in  Fig. \ref{fig: spectrum}b) where $\alpha_1=\infty$ and $\alpha_2 = 1.5$, and in Fig. \ref{fig: spectrum}c) where $\alpha_1=\alpha_2 = 1.5$. Although the mapping with the quantum Ising model is no longer exact for finite $\alpha_1$,$\alpha_2$, the two critical points are still referred to as ferromagnetic and antiferromagnetic. However, we notice that the location of the antiferromagnetic critical point is $\alpha_1$-dependent since $h_c = -1+2^{\alpha_1-1}$\,\cite{UhrichPRB2020}. Moreover, the spectrum dispersion relation in the proximity of the critical modes is affected by the presence of long-range interactions, thus leading to $\alpha$-dependent critical exponents\,\cite{UhrichPRB2020,DefenuArXiv2021}. Further details on the dispersion relations at the quantum critical points and for different values of $\alpha_1$ and $\alpha_2$ are provided in Appendix \ref{app: Taylor expansion}. 
	\subsection{Description of the cycle}
	\begin{figure}%[t!]
		\centering
		\includegraphics[width=0.5\textwidth]{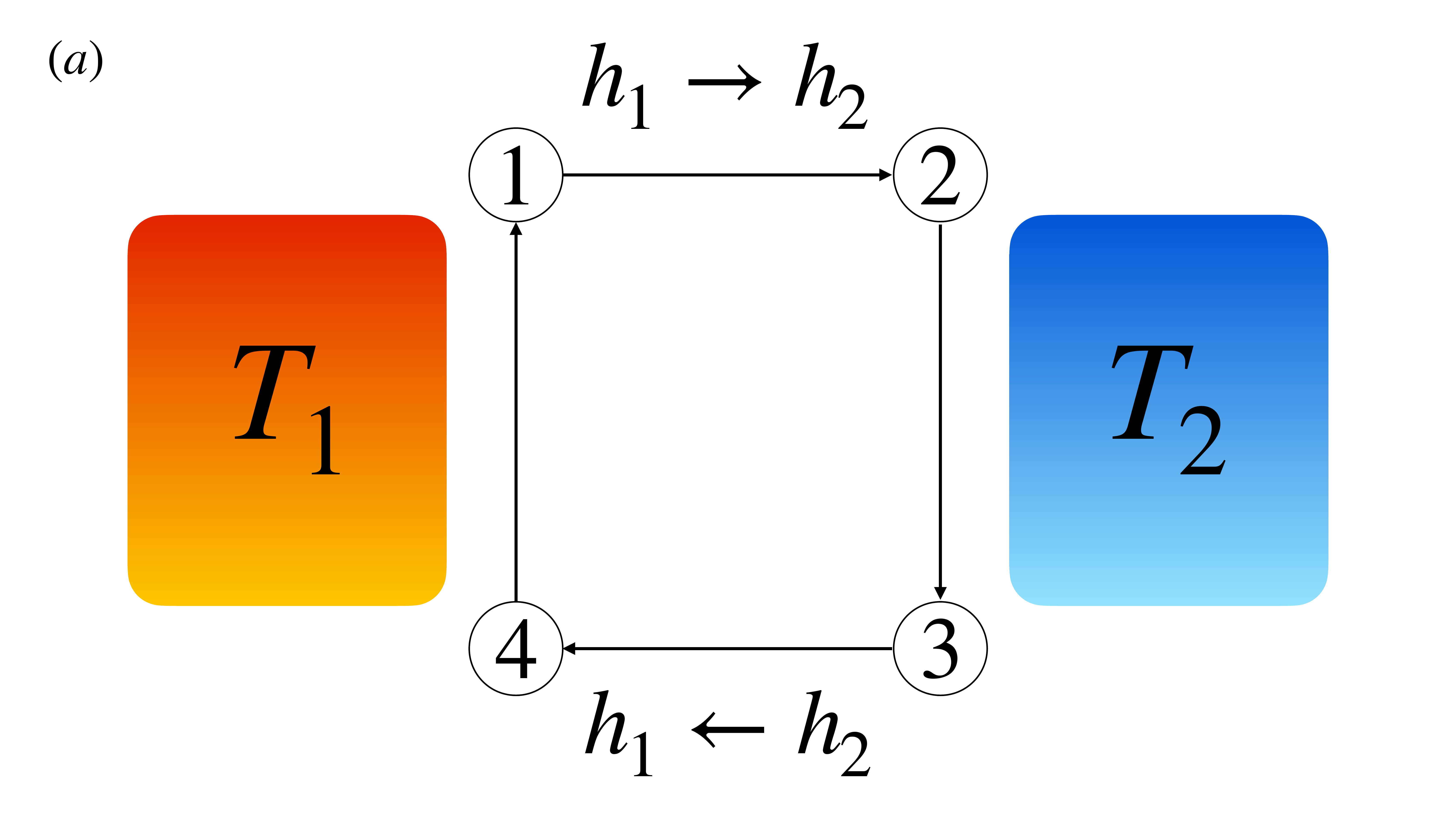}
		\centering
		\includegraphics[width=0.5\textwidth]{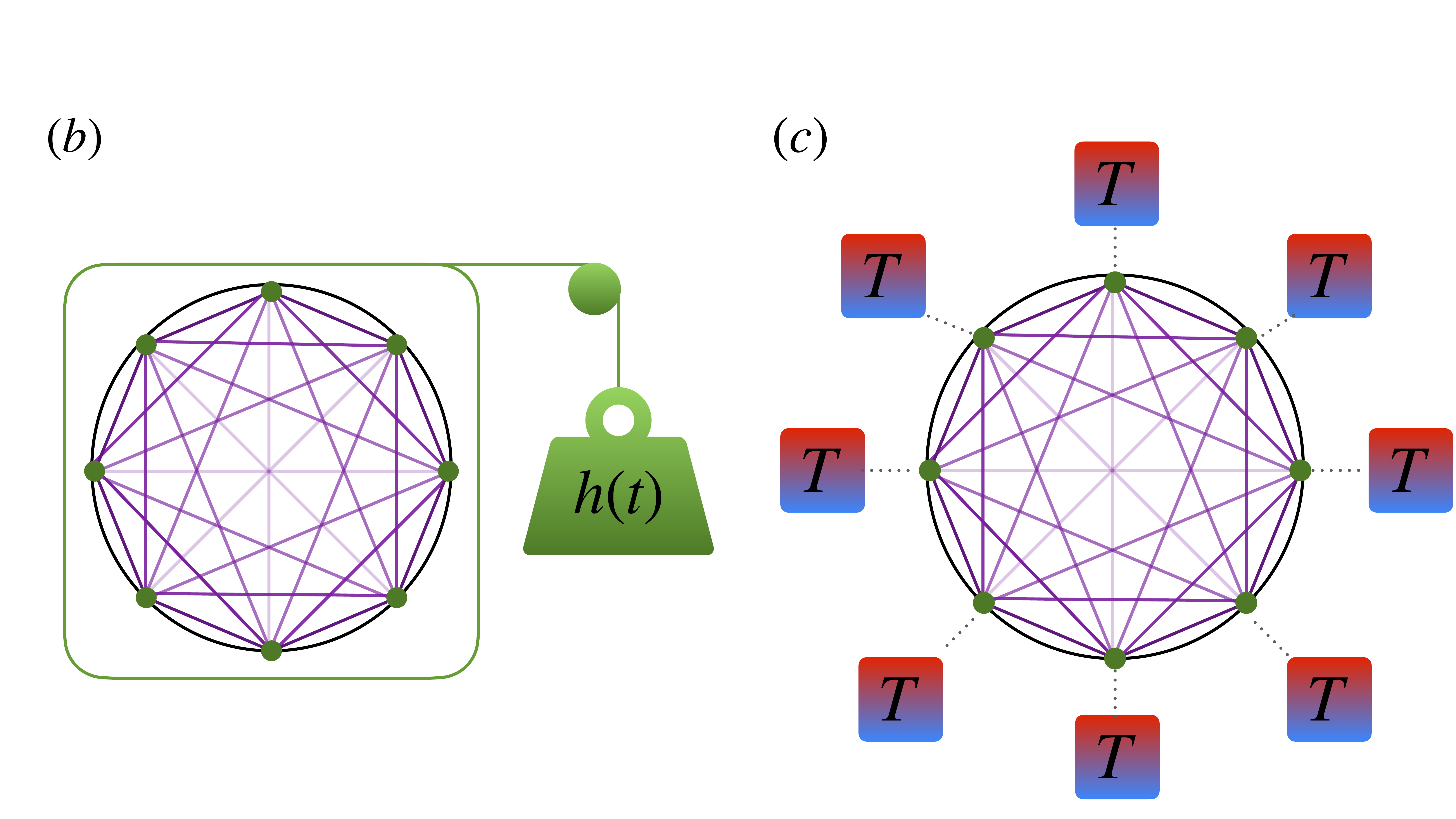}
		\caption{a) Schematic representation of the quantum Otto cycle. Sketch of the unitary b) and thermalization c) strokes for a many-body working substance with long-range interactions.}
		\label{fig: Ottocycle}
	\end{figure}
	%
	%
	%\begin{figure}%[t!]
	%	\centering
	%	\includegraphics[width=0.45\textwidth]{Strokes.pdf}
	%	\caption{Schematic representation of the quantum Otto cycle.}
	%	\label{fig: Strokes}
	%\end{figure}
	%
	The quantum Otto cycle\,\cite{KosloffEntropy2017,Deffner19Book,SolfanelliPRB2020}, consists of the following four strokes (see Fig. \ref{fig: Ottocycle}):
	\begin{itemize}
		\item Initially the system is assumed to be in thermal equilibrium with the hot reservoir at temperature $T_1 = 1/\beta_1$ and $h = h_1$. Then, in the first stroke (unitary decrease of $h$, $1\to 2$) the system is decoupled from the bath and it undergoes a unitary evolution where the chemical potential changes from $h_1$ to $h_2$.
		\item In the second stroke (thermalization at fixed $h$, $2\to 3$) the chemical potential is kept fixed at $h = h_2$, while the system is coupled with the thermal bath 2 so as to reach equilibrium at the temperature $T_2 = 1/\beta_2$. 
		\item In the third stroke (unitary increase of $h$, $3\to 4$) the system undergoes
		another unitary driving which brings the chemical potential back to the initial value $h_2\to h_1$.
		\item In the fourth and last
		stroke (thermalization at fixed $h$, $4\to 1$) the system at fixed $h=h_1$ is again coupled with bath 1 so as to reach the equilibrium at temperature $T_1$, thus closing the thermodynamic cycle. 
	\end{itemize}
	At the beginning of each stroke of the cycle (points $1$, $2$, $3$, $4$ in Fig.\ref{fig: Ottocycle}), the state of the system and the corresponding average energy are given by
	
	\begin{subequations}
		\begin{align}
			&\rho_1 = e^{-\beta_1H_1}/Z_1, &E_1 = \mathrm{Tr}\rho_1H_1,\label{eq: rho1}\\
			&\rho_2 = U\rho_1 U^\dagger, &E_2 = \mathrm{Tr}\rho_2H_2,\label{eq: rho2}\\
			&\rho_3 = e^{-\beta_2H_2}/Z_2, &E_3 = \mathrm{Tr}\rho_3H_2,\label{eq: rho3}\\
			&\rho_4 = \tilde{U}\rho_3 \tilde{U}^\dagger, &E_4 = \mathrm{Tr}\rho_4H_1,\label{eq: rho4}
		\end{align}
	\end{subequations}
	where $H_i = H(h_i)$, $Z_i = \mathrm{Tr}e^{-\beta_iH_i}$ $i=1,2$, $U$ and $\tilde{U}$  are the unitary evolution operators associated to the first and the third stroke respectively. In the following, we shall assume a linear time dependence of the chemical potential during the unitary strokes. Then the driving protocol corresponding to the first step $1\rightarrow 2$ can be written as
	\begin{align}
		h(t) = h_1 -\delta t\quad \mathrm{for}\quad t\in [0,\tau],\label{eq: driving protocol}
	\end{align}
	where $\delta = (h_1-h_2)/\tau$ is the sweep rate. The driving protocol during the third step of the cycle $3\rightarrow 4$ is given by $\tilde{h}(t) = h(\tau-t)$, for $t\in [0,\tau]$. The corresponding unitary evolutions then read 
	\begin{align}
		U &= \mathrm{T}\exp\left[-i\int_0^\tau dtH[h(t)]\right],\\
		\tilde{U} &= \mathrm{T}\exp\left[-i\int_0^\tau dtH[\tilde{h}(t)]\right],
	\end{align}
	where $\mathrm{T}\exp$ denotes the time-ordered exponential. During the second and the fourth strokes the external driving is switched off and the system interacts only with the baths, reaching thermal equilibrium. While long-range interacting systems are known to evade thermalization allowing for quasistationary states\,\cite{AntoniPRE1995,DefenuPNAS2021}, it can be shown that thermal equilibrium is safely reachable when $\alpha_1,\alpha_2>1$. The behavior of the long-range Kitaev chain coupled to thermal reservoirs has been recently addressed in Ref.\,\,\cite{KingArXiv2022}. 
	In particular, one can couple the system to $N$ thermal baths, one for each lattice site, all characterized by the same temperature $T$. The internal baths dynamics are described by a continuous model of free fermions with Hamiltonian\,\cite{D'Abbruzzo21PRB104}
	\begin{align}
		H_B = \sum_{n}\int dq \epsilon_n(q)b^\dagger_n(q)b_n(q),
	\end{align} 
	which is coupled to the chain through the linear interaction Hamiltonian
	\begin{align}
		H_{SB} = \sum_{n}\int dq g(q)(c_n+c_n^\dagger)(b_n(q)+b_n^\dagger(q)).
	\end{align}
	Assuming that the Born-Markov and the secular approximations\,\cite{Breuer02} are satisfied, then a Linblad master equation can be derived\,\cite{D'Abbruzzo21PRA103} for the system density operator $\rho$, reading 
	\begin{align}
		\frac{d\rho}{dt} = -i[H+H_{Ls},\rho]+\mathcal{D}[\rho],
		\label{eq: Lindblad master equation}
	\end{align}
	where $H_{Ls}$ is the Lamb-shift correction to the system Hamiltonian, and the dissipator $\mathcal{D}$ has the form\,\cite{D'Abbruzzo21PRB104}
	\begin{align}
		\mathcal{D}[\rho] = &\sum_k\mathcal{J}(\omega_k)[(1-f(\omega_k))(2\gamma_k\rho\gamma_k^\dagger-\{\gamma_k^\dagger\gamma_k,\rho\})\notag
		\\
		&+f(\omega_k)(2\gamma_k^\dagger\rho\gamma_k-\{\gamma_k\gamma_k^\dagger,\rho\})],
	\end{align}
	$f(\omega_k) = (1+e^{\beta\omega_k})^{-1}$ being the Fermi-Dirac distribution, and $\{A,B\} = AB+BA$ representing the anticommutator between the $A$ and $B$ operators, and $\mathcal{J}(\omega) = \pi\int dq|g(q)|^2\delta(\omega-\epsilon(q))$ the bath spectral density. The solution of the Lindblad master equation~\eqref{eq: Lindblad master equation} shows that the populations of the various normal modes evolve independently from one another towards the steady state, each with a decay rate equal to $r_k = 2\mathcal{J}(\omega_k)$\,\cite{KingArXiv2022} When the latter is nonzero for all $k$, the steady state is unique and is characterized by the thermal expectation values\,\cite{D'Abbruzzo21PRA103}
	\begin{align}
		\langle\gamma_k^\dagger\gamma_k\rangle = f(\omega_k) = \frac{1}{1+e^{\beta\omega_k}}.\label{eq: equilibrium populations}
	\end{align} 
	This justifies the use of the canonical thermal equilibrium state $\rho = e^{-\beta H}/Z$, with $Z = \mathrm{Tr}[e^{-\beta H}]$, in Eqs.~\eqref{eq: rho1} and~\eqref{eq: rho3}. Then, all the thermodynamic quantities can be easily computed using Eq.~\eqref{eq: equilibrium populations} or, equivalently, directly computing the partition function $Z$. In particular the internal energy reads
	\begin{align}
		E_i &= \sum_k\omega_{k,i}\left(\langle\gamma_k^\dagger\gamma_k\rangle-1/2\right)\notag
		\\
		&=-\sum_{k>0}\omega_{k,i}\tanh\left(\frac{\beta\omega_{k,i}}{2}\right),
	\end{align}
	with $\omega_{i,k} = \omega_k(h_i)$, for $i = 1,2$.
	Since during the second and
	the fourth strokes the external driving is switched off and the system interacts only with the baths, then energy is exchanged with them only in the form of heat
	\begin{align}
		&Q_1 = E_1-E_4\\
		&Q_2 = E_3-E_2.
	\end{align}
	According to the first law of thermodynamics we also have
	\begin{align}
		W = Q_1+Q_2.
	\end{align} 
	The three average energy exchanges $Q_1$,$Q_2$ and $W$ completely characterize the cycle operation.
	\subsection{Operation modes}
	Depending on the signs of $Q_1$, $Q_2$ and $W$, our engine may operate in any of the following four modes
	\begin{subequations}
		\begin{align}
			&[E]: Q_1\geq 0, Q_2\leq 0, W\geq 0;\label{eq:E}\\
			&[R]: Q_1\leq 0, Q_2\geq 0, W\leq 0;\label{eq:R}\\
			&[A]: Q_1\geq 0, Q_2\leq 0, W\leq 0;\label{eq:A}\\
			&[H]: Q_1\leq 0, Q_2\leq 0, W\leq 0;\label{eq:H}
		\end{align}
		\label{eq: operation modes}\end{subequations} 
	where $[E]$ denotes energy extraction (heat engine), $[R]$ denotes refrigerator, $[A]$ denotes thermal accelerator, and $[H]$ denotes heater\,\cite{Buffoni19PRL122}.
	The Kitaev chain can be decomposed into a collection of $N$ non-interacting qubits, each with level spacing $\omega_k$, (see Eqs.~\eqref{eq:Hdiag} and~\eqref{eq: spectrum}), which act as independent quantum thermal machines. In fact, the energy exchanges of the many-body device can be written as
	\begin{align}
		Q_1 =\sum_{k}Q_{1,k},\quad 
		Q_2 =\sum_{k}Q_{2,k},\quad
		W = \sum_{k}W_{k},
	\end{align}
	where $Q_{1,k}$, $Q_{2,k}$ and $W_k$ denote the energy flows corresponding to the $k$-th mode. Let us introduce the transition probabilities $1-P_k$, i.e., the probability of a nonadiabatic transition among the levels of the $k$-th qubit during the unitary stroke of the cycle. Then the energy exchanges take the form
		\begin{subequations}
			\begin{align}
				&Q_1 =-\sum_{k>0}\omega_{1,k}\left(f_{1,k}+f_{2,k}(1-2P_k)\right),\label{eq: Q1 nonadiabatic}\\
				&Q_2 =-\sum_{k>0}\omega_{2,k}\left(f_{2,k}+f_{1,k}(1-2P_k)\right),\label{eq: Q2 nonadiabatic}\\
				&W = Q_1+Q_2, \label{eq: W nonadiabatic}
			\end{align}\label{eq: nonadiabatic exchanges}\end{subequations}
	where we have introduced the shortcut notation $f_{i,k} = \tanh(\beta_i\omega_{i,k}/2)$, for $i= 1,2$.
	Let us notice that, as the system undergoes an Otto cycle, different qubits may operate in different regimes, among the $[E]$, $[R]$, $[A]$, or $[H]$ mode. As a consequence, the resulting operation mode of the many-body engine depends non-trivially on the interplay between the different qubit engines. 
	
	%A rough indication of the thermodynamical quantities is given by the Einstein approximation, in which the chain is replaced by $N$ identical single qubits with frequency
	%\begin{align}
	%\bar{\omega} = \frac{1}{N}\sum_k\omega_k,\label{eq: average spacing}
	%\end{align}
	%i.e., the average level spacing. [Espandi]

	%Then, when all the two level engines operate in the same operation, this also corresponds to the operation mode of the many body engine. On the other hand even if the complete setup works in a given operation mode, say heat engine $[E]$, the individual fermionic modes may act as, refrigerator, or thermal accelerator, depending on the details of the parameters characterizing the cycle.
	
	\section{Adiabatic cycle}\label{sec: adiabatic cycle}
\begin{figure*}
		\centering
		\includegraphics[width=\textwidth]{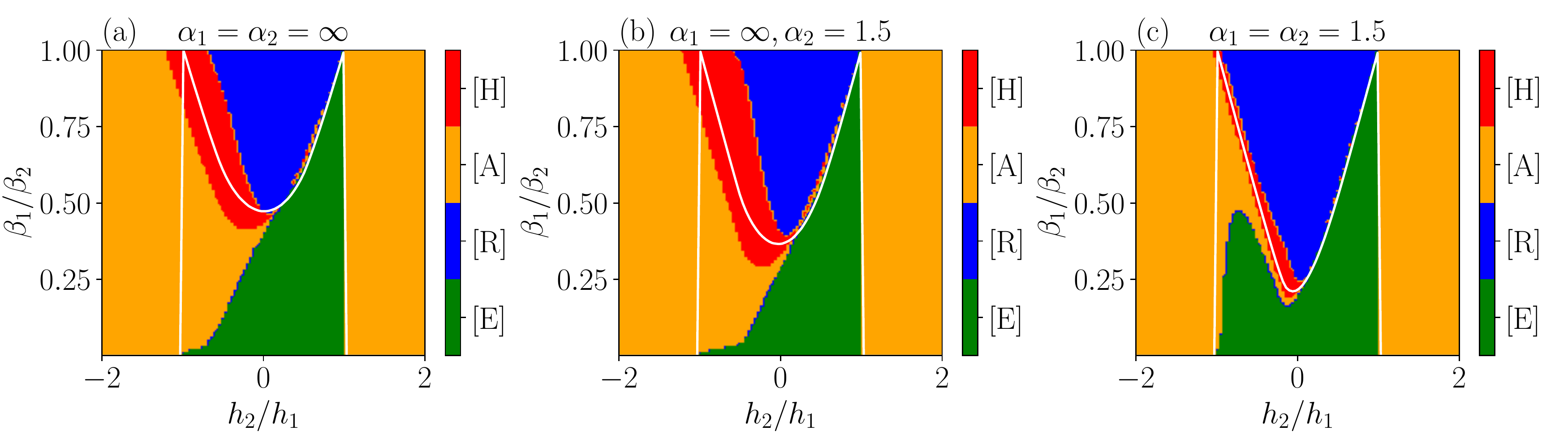}
		\caption{Operation mode diagram. Regions in the space of the parameters $\beta_1/\beta_2$ and $h_2/h_1$, corresponding to different operation modes. Different colors indicate different operation modes: blue indicates the refrigerator $[R]$, green indicates the heat engine $[E]$, yellow stays for the thermal accelerator $[A]$, and red for the heater $[H]$. White lines indicate the boundaries of the operation mode regions obtained in the Einstein approximation~\eqref{eq:single qubit operation modes}. The system size is fixed to $N = 200$, while the values of the initial temperature and chemical potential are chosen to be $T_1 = 100$, $h_1 = 2$.}
		\label{fig: Regimes}
	\end{figure*}

	Let us now analyze the case of an infinitely slow cycle i.e., the limit $\delta\to 0$ ($\tau\to\infty$). This regime is usually referred to as adiabatic, since the unitary evolution is sufficiently slow for the adiabatic theorem to hold, preventing transitions between the instantaneous eigenstates of the Hamiltonian, and leading to $P_{k}\sim1-O(\tau^{-2})$. The adiabatic approximation is known to break down as the energy gap closes. Strictly speaking, however, this happens only in the thermodynamic limit $N\to\infty$. Accordingly, for any finite $N$, one can choose the driving time scale such that the adiabatic approximation is justified, allowing us to set $P_k = 1$ in Eqs.~\eqref{eq: nonadiabatic exchanges} obtaining 
	\begin{subequations}
		\begin{align}
			&Q_1 =\sum_{k>0}\omega_{1,k}\left(f_{2,k}-f_{1,k}\right),\label{eq: Q1 adiabatic}\\
			&Q_2 =-\sum_{k>0}\omega_{2,k}\left(f_{2,k}-f_{1,k}\right),\label{eq: Q2 adiabatic}\\
			&W =\sum_{k>0}(\omega_{1,k}-\omega_{2,k})\left(f_{2,k}-f_{1,k}\right).\label{eq: W adiabatic}
		\end{align}
	\end{subequations}
	Figure \ref{fig: Regimes}  shows the regions of the parameters, $\beta_2$ and $h_2$, for fixed values of $\beta_1$ and $h_1$, corresponding to the different operation modes~\eqref{eq: operation modes}. These are compared with corresponding regions in the Einstein approximation, in which the chain is replaced by $N$ identical single qubits with frequency
	\begin{align}
		\bar{\omega} = \frac{1}{N}\sum_k\omega_k,\label{eq: average spacing}
	\end{align}
	i.e., the average level spacing. The region boundaries obtained in this approximation correspond to the white lines in Fig. \ref{fig: Regimes} and provide a rough estimation for the engine operation mode. Since, in this limit, only one level-spacing is present, one can obtain these boundaries by applying the results of Ref.\,\,\cite{SolfanelliPRB2020} for the operation modes of a single qubit, namely
	\begin{subequations}
		\begin{align}
			&[E]: \frac{\beta_1}{\beta_2}\leq\frac{\bar{\omega}_{2}}{\bar{\omega}_{1}}\leq 1,\\
			&[R]: \frac{\bar{\omega}_{2}}{\bar{\omega}_{1}}\leq\frac{\beta_1}{\beta_2},\\
			&[A]: \frac{\bar{\omega}_{2}}{\bar{\omega}_{1}}\geq 1 
		\end{align}\label{eq:single qubit operation modes}\end{subequations}where $\bar{\omega}_{1,2}$ corresponds to Eq.~\eqref{eq: average spacing} for $h=h_{1,2}$ respectively and we have assumed, without loss of generality, $\beta_1\leq\beta_2$.
	Let us notice that, as conditions~\eqref{eq:single qubit operation modes} rule out the possibility of a single qubit acting as a heater ($[H]$), this regime cannot be well described within the mean-spacing approximation in the adiabatic limit.
	
	In the region, $h_2/h_1>0$, however, in which the heater phase is not present, the operation mode phase diagram is well reproduced by the Einstein approximation, see Eq.~\eqref{eq:single qubit operation modes}, regardless of the values of $\alpha_{1}$ and $\alpha_{2}$. By comparing the three diagrams in Fig.\,\ref{fig: Regimes} we notice that the case $\alpha_1 = \infty$, first studied in Ref.\,\cite{Wang20PRE102}, is pretty similar to the nearest-neighbours limit, and does not lead to substantial advantage. On the other hand, in the presence of both long-range hopping and long-range pairing ($\alpha_1 = \alpha_2 = 1.5$) the heater region is greatly reduced, and the Einstein approximation works better even in the $h_2/h_1 <0$ region.
	
	In conclusion, the presence of long-range interactions may generate a substantial advantage provided proper values of the decay exponents $\alpha_{1}$ and $\alpha_{2}$ are chosen. Indeed, in the equally long-range case (Fig.\,\ref{fig: Regimes}c) the $[R]$ and $[E]$ regimes, which are the most relevant to technological applications, are enhanced with respect to the nearest neighbor case (Fig.\,\ref{fig: Regimes}a)), and become prevalent in the whole parameter region $|h_2/h_1| < 1$ . In our studies, the most profitable configuration of the Hamiltonian in Eq.~\eqref{eq: long-range Kitaev chain H} corresponds to the realistic case, where the Kitaev Hamiltonian approximately represents the long-range Ising model, i.e.,  $\alpha_1 = \alpha_2$, which will be the focus of the following discussion.
	
	\subsection{Heat Engine operation mode}\label{sec: Heat Engine operation}
	\begin{figure}
		\centering
		\includegraphics[width=0.8\linewidth]{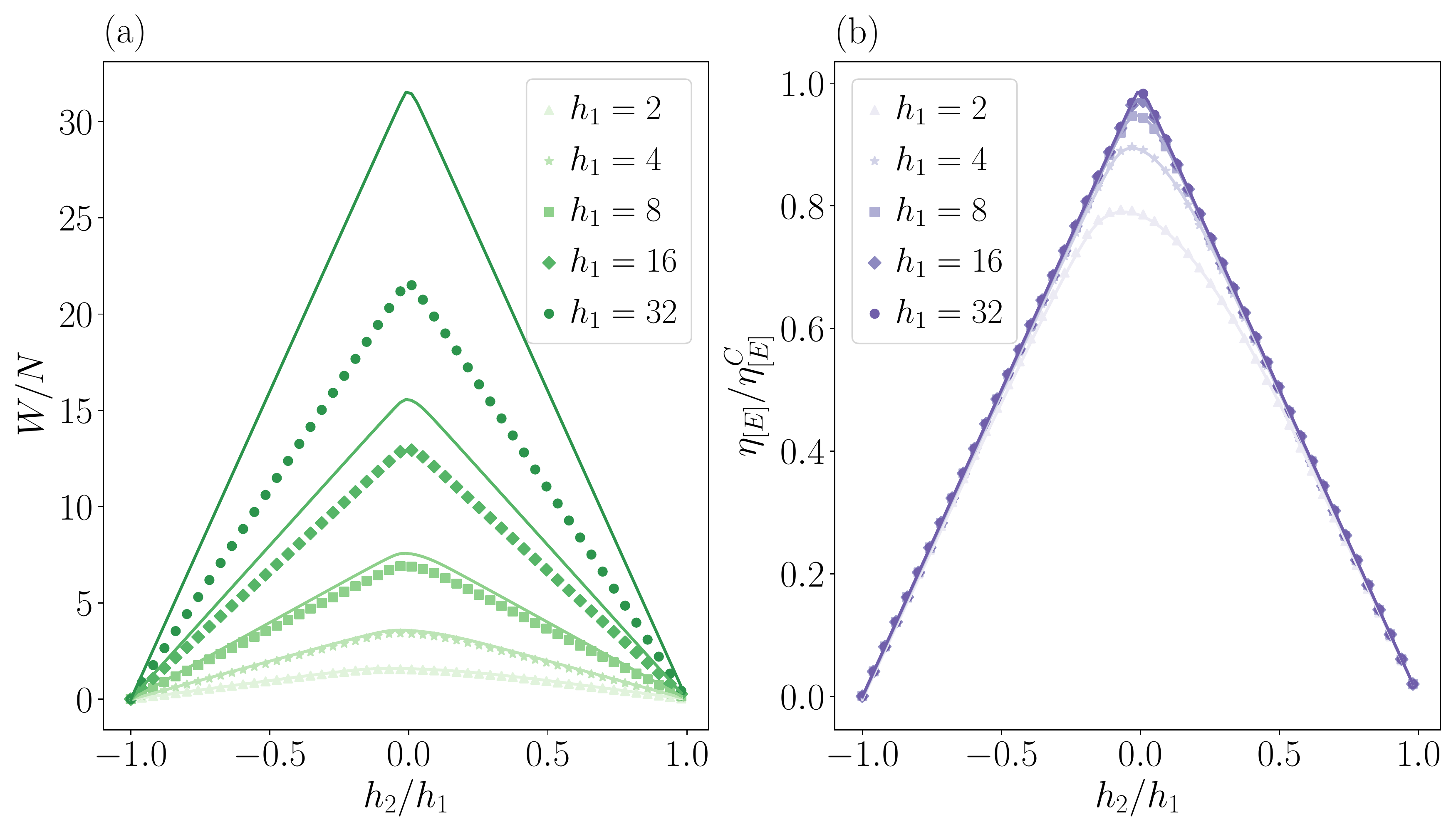}
		\centering
		\caption{Work output (panel a) and engine efficiency (panel b), plotted as a function of $h_2/h_1$, for different values of $h_1$ corresponding to different colors. Exact values~\eqref{eq: W adiabatic} are represented as scatter plots with different markers (one for each value of $h_1$), while bold lines refer to the approximated result of Eq.~\eqref{eq: W0}. The system size is $N = 200$, the temperatures of the baths are fixed to $T_1 = 100$, $T_2 = 0.01$.}
		\label{fig: W_eta_h1}
	\end{figure}
	\begin{figure}
		\centering
		\includegraphics[width=0.8\linewidth]{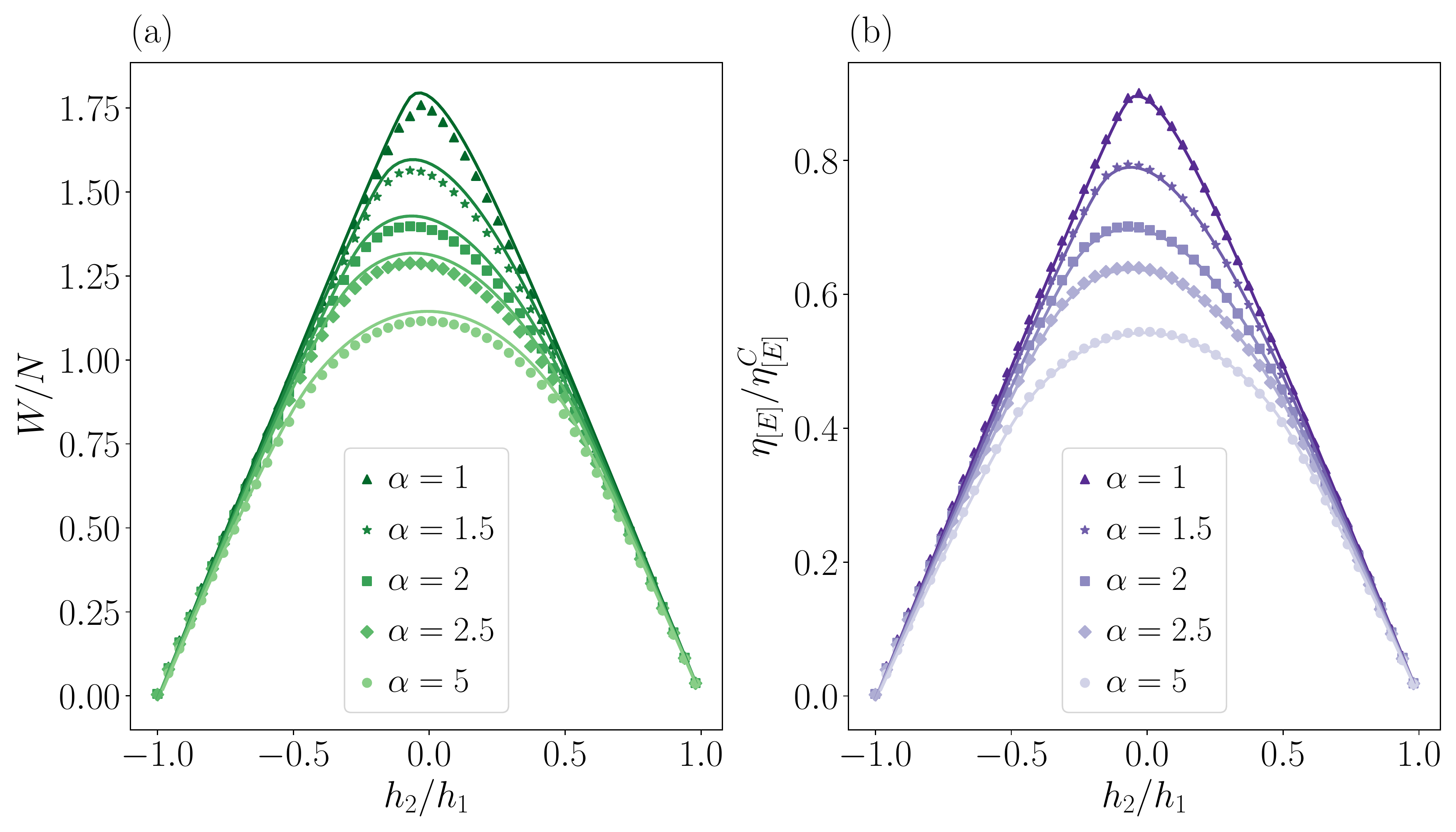}
		\centering
		\caption{Work output (panel a) and engine efficiency (panel b), plotted as a function of $h_2/h_1$, for different values of $\alpha_1 = \alpha_2 = \alpha$ corresponding to different colors. Exact values~\eqref{eq: W adiabatic} are represented as scatter plots with different markers (one for each value of $\alpha$), while bold lines to the approximate result of Eq.~\eqref{eq: W0}. The system size is $N = 200$, the temperatures of the baths are fixed to $T_1 = 100$, $T_2 = 0.01$ and the initial chemical potential value is $h_1 = 2$.}
		\label{fig: W_eta_a}
	\end{figure}
	%
	%The purpose of a heat engine is to exploit the energy flow from a hot reservoir to a cold one to generate work. 
	The purpose of a heat engine is to exploit natural flow of heat from a hot reservoir to a cold one to generate work. Thus, the performance of a device operating in the $[E]$ mode may be optimized by maximizing the work output. %A second possibility is to maximize 
	Another estimator of the engine performance is the heat engine efficiency, defined as the ratio between the energy gain in the form of work and the heat extracted from the hot reservoir
	\begin{align}
		\eta_{[E]} = \frac{W}{Q_1} = 1+\frac{Q_2}{Q_1}.
	\end{align}
	The second law of thermodynamics imposes this efficiency to be always smaller than the Carnot efficiency $\eta_{[E]}^C = 1-T_2/T_1$\,\cite{Fermi1956}.
	The functioning of a heat engine is naturally boosted when the difference between the temperatures of the baths is large, in fact in this situation the energy flow and consequently also the work extraction, are favored. Indeed, in this regime, the Carnot efficiency gets close to unity. This basic physical intuition leads us to consider the region of parameters where $T_2\ll T_1$ as the most interesting for the $[E]$ operation. More precisely, we take $T_2 \ll \bar{\omega}(h) \ll T_1$, with $\bar{\omega} (h)$ playing the role of a typical energy scale of the system. Accordingly, %More precisely we can introduce the energy scale  $\mathcal{E}$ of the Kitaev chain working substance, as $\mathcal{E}\sim N^{-z}$ for $N\gg |h-h_c|^{-\nu}$ and $\mathcal{E}\sim |h-h_c|^{\nu z}$ for $N\ll |h-h_c|^{-\nu}$\,\cite{Sachdev}, where $z$ and $\nu$ are the dynamical and correlation length critical exponents, respectively. Then we consider a cold bath with temperature $T_2\ll \mathcal{E}$ and a hot bath with temperature $T_1\gg \mathcal{E}$. 
	the working substance is close to the ground state when it is in equilibrium with the cold bath, and it is close to the maximally mixed state when in equilibrium with the hot bath. As a consequence, the work extracted from the working substance can be written as
	\begin{align}
		W \simeq W_0 \equiv N(\bar{\omega}_1-\bar{\omega}_2)/2,\label{eq: W0}
	\end{align}
		(see Appendix \ref{app: Work output in the infinite temperature gradient limit} for details) which is fully determined by the average level spacing $\bar{\omega}_i$, $i=1,2$ for $h=h_{1,2}$. It follows that the optimal work output is reached for for the values of $h_1$, $h_2$ that respectively maximise and minimise the function $\bar{\omega}(h)$ in Eq. \ref{eq: average spacing}, namely:
	\begin{align}
		W_{\mathrm{max}} \simeq W_{0,\mathrm{max}} =  \frac{N}{2}(\max_{h}[\bar{\omega}]-\min_{h}[\bar{\omega}]),
	\end{align}
	where the optimization has to be performed over the values of $h$ compatible with the approximation~\eqref{eq: W0}, i.e., such that $T_2 \ll \bar{\omega}(h) \ll T_1$. %\nic{we denote as $\mathcal{A}$ the region of parameters $h$, which satisfy this condition}.
	Within the same approximation, the heat engine efficiency reads
	\begin{align}
		\eta_{[E]} \simeq \eta_{[E]}^0\equiv 1-\frac{\bar{\omega}_2}{\bar{\omega}_1}\leq1-\frac{\min_{h}[\bar{\omega}]}{\max_{h}[\bar{\omega}]}
	\end{align}
	
	Remarkably, this choice of $h_1$ and $h_2$ allows to maximize both the work output and the cycle efficiency. In Fig. \ref{fig: W_eta_h1}a) and b) the exact work output $W$, and the exact engine efficiency $\eta_{[E]}$ respectively (dotted lines) are compared with $W_0$ and $\eta^0_{[E]}$ (solid lines) for different values of $h_1$, $h_2$ and with $T_1 = 100, T_2=0.01$ and $\alpha_1 = \alpha_2 = 1.5$. We notice that the approximation $W \simeq W_0$ breaks down for large values of $h_1$, when $\bar{\omega} (h_1)$ becomes of the same order of $T_1$, while $\eta^0_{[E]}$ remains a good estimate of $\eta_{[E]}$ even in this regime. Finally, let us notice that, regardless of the validity of the approximation, the maxima of $\eta_{[E]}$ and $W$ are actually close.
	
\begin{figure}
		\centering
		\includegraphics[width=0.5\linewidth]{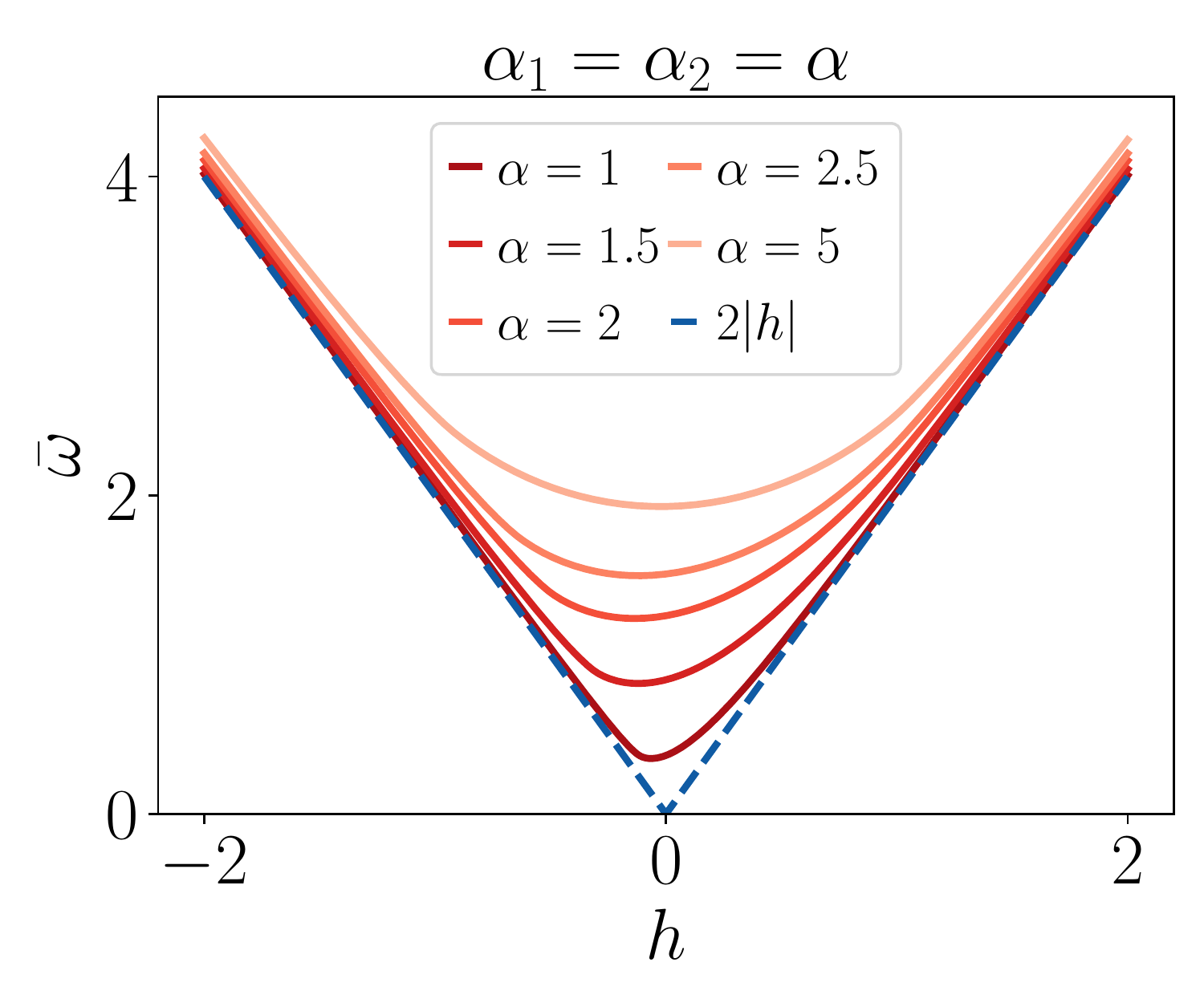}
		\centering
		\caption{Average level spacing~\eqref{eq: average spacing}, as a function of the chemical potential $h$. Different colors correspond to different values of $\alpha_1 = \alpha_2 = \alpha$. The blue dashed line represents the asymptotic value in the large $|h|$ limit $\bar{\omega}\approx 2|h|$. The system size is $N = 500$. The system size is $N = 500$.}
		\label{fig: average_spectrum}
	\end{figure}

	In Fig. \ref{fig: W_eta_a} we plot $W$ and $\eta_{[E]}$ against $h_2/h_1$, for different values of $\alpha$, showing that they grow as the range of the interaction increases, signaling a clear advantage of the long-range regime. This advantage can be traced back to the properties of the spectrum of the system, encoded in the average level spacing $\bar{\omega}$. In fact, the minimum of $\bar{\omega}(h_2)$, which corresponds to the maximum of both $W_0$ and $\eta^0_{[E]}$, is affected by the presence of long-range interactions as shown in Fig. \ref{fig: average_spectrum}.
	\subsection{Refrigerator operation mode}\label{sec: Refrigerator operation mode}
	\begin{figure}%[t!]
		\centering
		\includegraphics[width=0.5\linewidth]{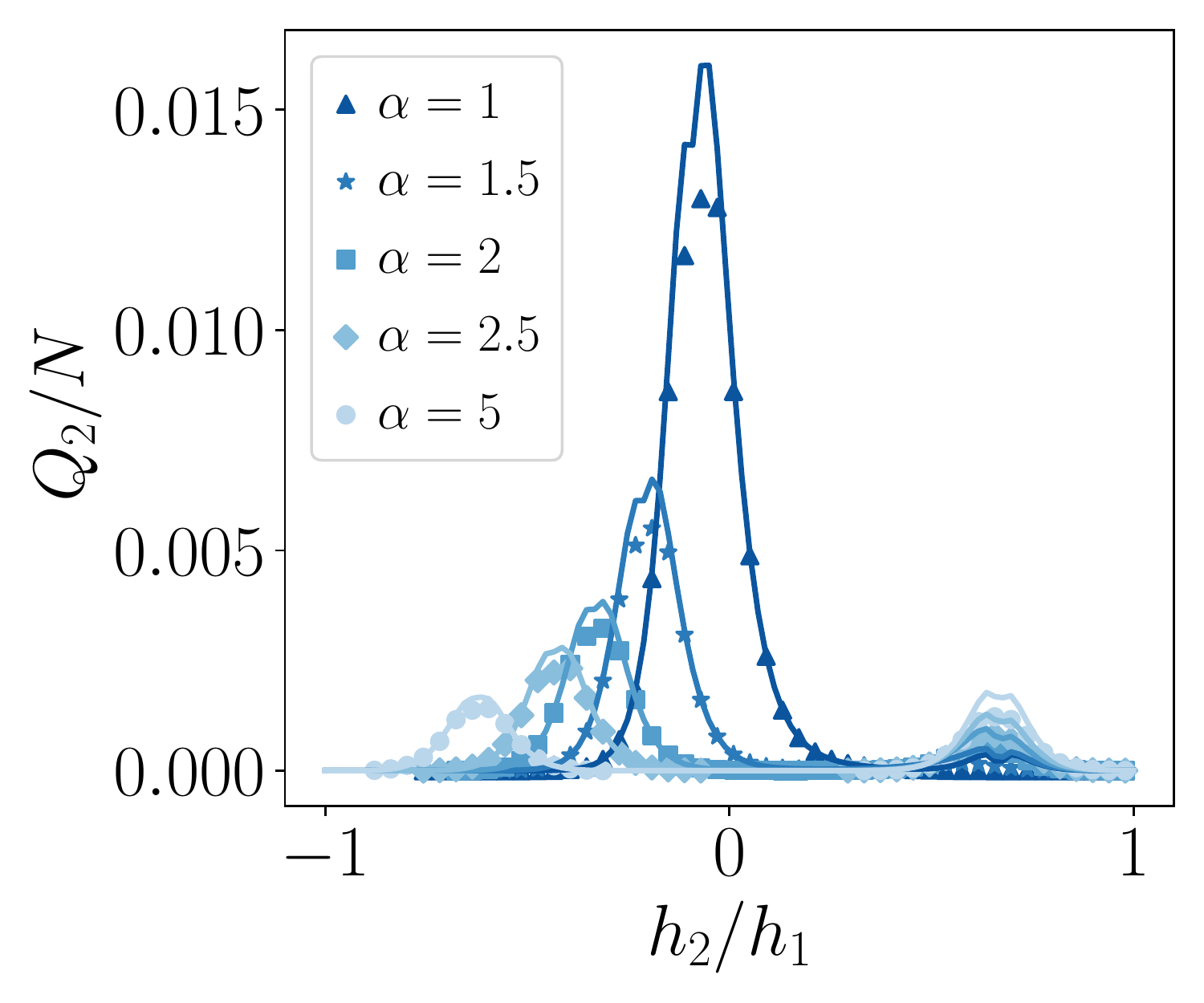}
		\caption{Heat extracted from the cold reservoir $Q_2$, as a function of $h_2/h_1$. Scatter plots indicate the exact values~\eqref{eq: Q2 adiabatic}, bold lines indicate the approximated result ~\eqref{eq: Q2 approx}. Different colors correspond to different values of $\alpha_1 = \alpha_2 = \alpha$. The baths temperatures are fixed to $T_1 = 0.1$, $T_2 \simeq 0.099$.}
		\label{fig: Q2_a10}
	\end{figure}
	%
	%
	%
	%The typical situation in which a quantum refrigerator operates to cool down a quantum system is characterized by two conditions on the two reservoirs temperatures $T_1$ and $T_2$. First of all 
	In the typical situation in which a quantum refrigerator operates we may expect the two temperatures to be pretty similar $T_2\lesssim T_1$, since we can imagine that also the baths are embedded in the same quantum hardware of the working substance. Moreover, we want the system to be deeply in the quantum regime, accordingly, the temperatures involved in the cycle should be small with respect to the system energy scale, $T_2\lesssim T_1\ll \bar{\omega}$. Under these assumptions, the heat extracted from the cold reservoir reads
	\begin{align}
		Q_2 \simeq \sum_{k>0}\omega_{2,k}e^{-\beta_2\omega_{2,k}}\left[1-\tanh\left(\frac{\beta_2\omega_{2,k}-\beta_1\omega_{1,k}}{2}\right)\right].\label{eq: Q2 approx}
	\end{align}
	Since the above expression is positive definite, we can conclude that within the considered approximation the Otto cycle always operates as a refrigerator. On the other hand, far from the quantum critical points
	\begin{align}
		Q_2\simeq N\min_{k}[\omega_{2,k}]e^{-\beta_2\min_{k}[\omega_{2,k}]}.\label{eq: Q2 exp decay}
	\end{align} 
	so that $Q_2$ has an exponentially decaying behavior as $T_2\to 0$.
	As $h_2$ becomes close to $h_c(\alpha)$ the spectrum is no longer gapped, so the above considerations do not apply. In this regime, the main contribution to $Q_2$ comes from the soft modes, resulting in a power law decay in $T_2$
	\begin{align}
		Q_2\simeq NK(\alpha)T_2^{1+1/z},\label{eq: Q2 power law}
	\end{align}
	where $z$ is the dynamical exponent which, in general, depends on $\alpha$. In particular for the $h_2 = 1$ ferromagnetic critical point we have
	\begin{align}
		z = \begin{cases}
			\alpha-1\quad&\mathrm{for}\quad\alpha<2\\
			2\quad&\mathrm{for}\quad\alpha>2,
		\end{cases}\label{eq: dynamical exponent 1}
	\end{align}
	while for the $h_2 = -1+2^{\alpha-1}$ antiferromagnetic critical point,
	\begin{equation}
		z = 1.\label{eq: dynamical exponent 2}
	\end{equation} 
	Further details on the derivation of Eqs.~\eqref{eq: Q2 approx},~\eqref{eq: Q2 exp decay},~\eqref{eq: Q2 power law} are provided in Appendix \ref{app: Cooling capability}.
	Since close to the ferromagnetic critical point $1/z$ grows indefinitely as $\alpha \rightarrow 1$, the presence of long-range interactions does not result in any advantage at low temperatures. However, close to the antiferromagnetic case, $z$ does not depend on $\alpha$: in this case the factor $K(\alpha)$ can actually provide an advantage. That this is indeed the case, is confirmed by the data shown in Fig.\,\ref{fig: Q2_a10}, where $Q_2/N$ is plotted as a function of $h_2$ for different values of $\alpha$ and ($T_1 = 0.1$, $T_2 = 0.099$). The figure shows a clear advantage as the range of the interaction is increased. 
	
	Thus, the different low-temperature scalings of Eq.~\eqref{eq: Q2 power law} lead to the peaks in $Q_2$ at $h_2/h_1 = h_c(\alpha)/h_1$, corresponding to an enhanced cooling capability at quantum criticality, as shown in Fig. \ref{fig: Q2_a10}. This effect is augmented when long-range interactions are present, leading to larger and larger peaks as $\alpha\to 1$ and showing long-range advantage also in the most significant regime, i.e., the refrigerator operation. It is worth noting that, while the heat engine configuration is optimized by a long-range interacting machine operating close to the ferromagnetic critical point, the refrigerator operates optimally in the vicinity of the antiferromagnetic critical point.
	\section{Finite time cycle} \label{sec: Finite time cycle}
	
	While in previous sections we have focused on the thermal machines operating in the adiabatic regime, assuming an infinitely slow unitary driving of the working substance,i.e., setting $P_{k}=1$ in Eq.~\eqref{eq: W nonadiabatic}, we are now going to consider the effects of a finite time driving and allow the working substance to have dynamically generated defects during the Otto cycle, causing nonadiabatic energy losses. We provide evidence of long-range advantages also in the more realistic case of a finite-time heat engine and refrigerator, by computing the universal scaling of the nonadiabatic energy losses with the driving speed. In contrast with the nearest-neighbor case, such energy losses are universally suppressed when long-range interactions are present.
	
	\subsection{Exact solution of the non adiabatic dynamics and adiabatic perturbation theory}
	In general, the unitary evolution generated by $H(h(t))$ is such that it only mixes the states $|0_{k},0_{-k}\rangle$ and $|1_{k},1_{-k}\rangle$, for each value of $k$. As a consequence, the dynamics of the Kitaev chain can be exactly described by the $N$ independent %Schr\"odinger
	evolution equations, each restricted to the two-dimensional subspace associated with the corresponding $k$-mode\,\cite{DziarmagaAdvancesinPhysics2010}. These can be cast into a matrix evolution for the Bogolyubov coefficients $u_k$,$v_k$~\eqref{eq: bogoliubov coefficients},
	\begin{align}
		i\frac{d}{dt}\begin{pmatrix}
			u_k\\
			v_k\end{pmatrix} = 
		\mathcal{H}_k(t)
		\begin{pmatrix}
			u_k\\
			v_k
		\end{pmatrix},\label{eq: unitary evolution}
	\end{align} 
	where $\mathcal{H}_k$ is given by Eq.~\eqref{eq:Hk}, with $h = h(t)$. By means of the transformation  $t' = \Delta_k(t_k-h+\delta t)/\delta$, this is mapped onto a Landau-Zener-St{\"u}ckelberg-Majorana (LZSM) problem\,\cite{Landau1932,ZenerProcRSoc1932,Stueckelberg1932,Majorana1932,DziarmagaAdvancesinPhysics2010}:
	\begin{align}
		i\frac{d}{dt'}\begin{pmatrix}
			u_k\\
			v_k
		\end{pmatrix} = \begin{pmatrix}
			-\Omega_k t' &1\\
			1 & \Omega_k t'
		\end{pmatrix}\begin{pmatrix}
			u_k\\
			v_k
		\end{pmatrix},\label{eq: LZSM}
	\end{align}
	where $\Omega_k = \delta/\Delta_k^2$. The exact general solution of Eq.~\eqref{eq: LZSM} can be written in terms of Weber (or parabolic cylinder) D-functions $D_\nu(z)$, %and Gamma functions $\Gamma(z)$
	(see Ref.\,\cite{DziarmagaAdvancesinPhysics2010}), leading to 
	\begin{align}
		&v_k(t') = aD_{-s-1}(-iz)+bD_{-s-1}(iz),\label{eq: exact v}\\
		&u_k(t') = \left(\Omega_kt'-2i\frac{\partial}{\partial t'}\right)v_k(t'),\label{eq: exact u}
	\end{align}
	with $s = (4i\Omega_k)^{-1}$, $z = \sqrt{\Omega_k}t'e^{i\pi/4}$, and $a,b$ arbitrary complex parameters to be fixed by the initial conditions $u_k(t_i)$, $v_k(t_i)$. Accordingly, the solution of Eq.~\eqref{eq: unitary evolution} reads:
	\begin{align}
		&|\psi(t)\rangle = \prod_k|\psi_k(t)\rangle,\\
		&|\psi_k(t)\rangle = u_k(t)|0_{k},0_{-k}\rangle+v_k(t)|1_{k},1_{-k}\rangle,
	\end{align} 
	where $u_k(t) = u_k(t'(t)),v_k(t) = v_k(t'(t))$. 
	\\
	We can introduce the instantaneous eigenstates of the two-level Hamiltonians $\mathcal{H}_k(t)$ at time $t$, given by
	\begin{align}
		|\phi^{\pm}_k(t)\rangle = \bar{u}_k(h(t))|0_{k},0_{-k}\rangle \pm \bar{v}_k(h(t))|1_{k},1_{-k}\rangle,\label{eq: istantanesous eigenstates}
	\end{align}
	with $\bar{u}_k(h) = \cos(\theta_k(h)/2)$, $\bar{v}_k(h) = \sin(\theta_k(h)/2)$, where $\theta_k(h) = \arctan(\Delta_k/(h-t_k))$ is the Bogoliubov angle for a chemical potential $h=h(t)$. The probability for the system to be found in the instantaneous eigenstates during the evolution reads
	\begin{align}
		P_k(t) &= |\langle \phi_k^{\pm}(t)|\psi_k(t)\rangle|^2\notag\\
		&=|\bar{u}_k(h(t))u_k(t)+\bar{v}_k(h(t))v_k(t)|^2.
	\end{align}
	By inserting the expression for $u_k$, $v_k$ in Eqs. ~\eqref{eq: exact v},~\eqref{eq: exact u}, in the above expression, one obtains an analytical expression for $P_k(t)$\,\cite{VitanovPRA1996}. This exact solution, however, is rather cumbersome.  Considering the limit of a slow driving protocol $\delta\to 0$, with final time $\tau = |h_f-h_i|/\delta\to\infty$ allows for a simpler description that captures and better grasp the relevant physics involved in the dynamics. In this regime, the first non-trivial correction to the $P_k$ takes the celebrated LZSM  
	form  
	\begin{align}
		P_k \simeq 1-\exp\left(-\frac{\pi\Delta_k^2}{\delta}\right) + O(\delta^2). \label{eq: LZSM formula}
	\end{align}
	see Ref.\,\cite{DeGrandi2010} for its derivation using adiabatic perturbation theory. Although for finite $\Delta_k$ the $O(\delta^2)$ contributions is leading, as the transition point is crossed, the physics is dominated by the soft modes with small $\Delta_k$. As a consequence, in any relevant thermodynamical quantity, the $O(\delta^2)$ contribution in the r.h.s. of~\eqref{eq: LZSM formula} is negligible with respect to the non-analytic exponential one. 
	\subsection{Universal dynamical scaling of the nonadiabatic energy losses}
	%
	%
	%The exact treatment for the finite time dynamics introduced in the previous section allows us to compute 
	Knowing the probabilities $P_k$ is sufficient to compute the energy exchanges with the two reservoirs and with the external drivings, during our finite-time Otto cycle, in Eqs~\eqref{eq: nonadiabatic exchanges}.
	%\begin{subequations}
	%	\begin{align}
	%		&Q_1 =-\sum_{k>0}\omega_{1,k}\left(f_{2,k}+f_{1,k}(1-2P_k)\right),\label{eq: Q1 nonadiabatic}\\
	%		&Q_2 =-\sum_{k>0}\omega_{2,k}\left(f_{1,k}+f_{2,k}(1-2P_k)\right),\label{eq: Q2 nonadiabatic}\\
	%		&W = Q_1+Q_2, \label{eq: W nonadiabatic}
	%	\end{align}\label{eq: nonadiabatic exchanges}\end{subequations}
	%When $P_k = 1,\forall k$, we recover the adiabatic limit of section  \ref{sec: adiabatic cycle}. 
	Let us consider a single mode and the corresponding two-level system formed by $|0_k,0_{-k}\rangle$, $|1_k,1_{-k}\rangle$. It is known\,\cite{SolfanelliPRB2020} that, when $P_k < 1$, a region in parameter space corresponding to the heater $[H]$ appears and it becomes the only possible regime when $P_k\leq 1/2$, since in this case the energy exchanges become negative definite. This happens for the many-body Kitaev chain as well if the driving is so fast that $P_k < 1/2$ for all the values of $k$. For any finite-time driving, the presence of finite transition probabilities $1-P_k$ hinders the engine performance enhancing the irreversible character of the cycle. This can be explicitly proven by computing the so-called entropy production (as shown in Appendix \ref{app: Entropy production}).   
	
\begin{figure}%[t!]
		\centering
		\includegraphics[width=0.4\textwidth]{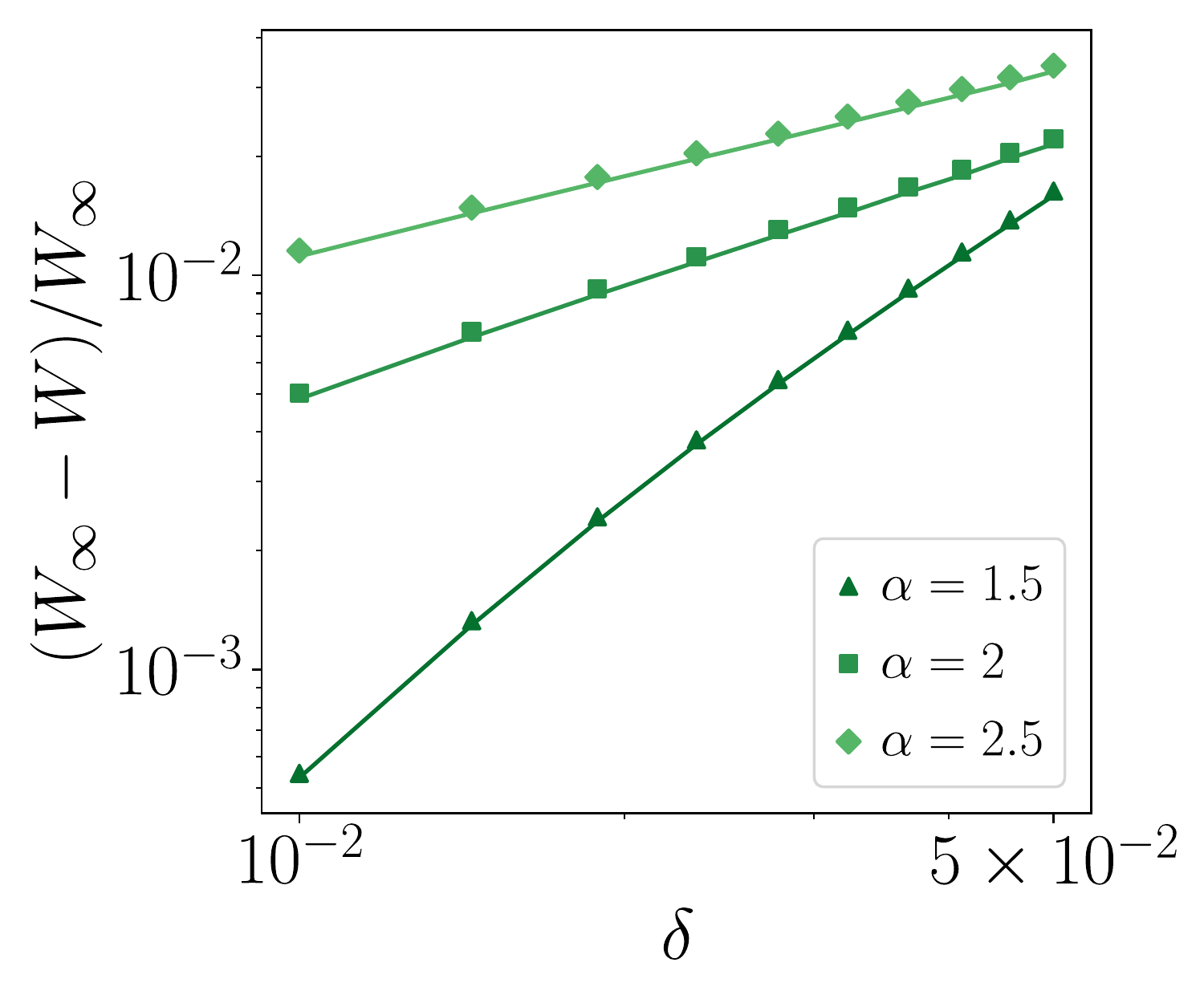}
		\caption{Nonadiabatic work loss ratio as a function of the driving velocity $\delta$, for different values of $\alpha_{1} = \alpha_{2} = \alpha$, corresponding to different colors and markers. Scatter plots indicate the exact numerical values while bold lines indicate the approximated result~\eqref{eq: W_infty-W optimal}. The cycle parameters $h = 2$, $h_2 = 0$, $T_1 = 100$, $T_2 = 0.01$. The system size is $N = 500$.}
		\label{fig: EnergyLoss}
	\end{figure}

	Here, we follow the ideas introduced in Ref.\,\cite{Revathy20PRR2}, about the dependence of the non-adiabatic energy losses from the velocity $\delta$ of a finite time cycle. Indeed, those can be directly related to the universal Kibble-Zureck scaling, i.e. the scaling of the density of excitations generated dynamically as the chemical potential slowly ramps across one of the critical points\,\cite{ZurekPRl2005}. Let us consider the heat engine operation in the optimal regime identified in section \ref{sec: Heat Engine operation} ( $T_2 \ll \bar{\omega} \ll T_1$). In this situation, the first unitary stroke cannot increase the entropy as the system already starts from the maximal entropy state $\rho_1\propto\mathbb{I}$. On the other hand, we expect defects to be generated in the fermionic chain during the third stroke of the cycle as the system is almost in its ground state $\rho_3\simeq |\mathrm{gs}\rangle\langle\mathrm{gs}|$ before the unitary evolution takes place. For a slow driving of the system through a quantum critical point, the density of excitations is given by
	\,\cite{DefenuPRB2019}
	\begin{align}
		n_{\mathrm{ex}} \equiv \frac{1}{N}\sum_{k}(1-P_k) \simeq \int_{-\pi}^\pi\frac{dk}{2\pi}e^{-\pi\Delta_k^2/\delta},\label{eq: excitation density}
	\end{align}
	where in the last step we used the result in Eq. \ref{eq: LZSM formula} and took the $N \rightarrow \infty$ limit. Since $\delta$ is small, the leading contribution to the integral comes from the soft modes with $\Delta_k \sim 0$. By considering the asymptotic scaling of the dispersion relation in correspondence of the ferromagnetic critical point $\Delta_k \sim |k|^{\min(\alpha_2-1,1)}$ (see Appendix \ref{app: Taylor expansion})  we find the scaling law\,\cite{DefenuPRB2019}
	\begin{align}
		n_{\mathrm{ex}}\sim \delta^\theta \quad\mathrm{with}\quad \theta = \begin{cases}
			(2\alpha_2-2)^{-1}\quad &\mathrm{for}\quad\alpha_2\leq2,\\
			1/2\quad &\mathrm{for}\quad\alpha_2>2.
		\end{cases}\label{eq: kibble_zurek_exponent}
	\end{align}
	%We notice that, for long-range regime pairing the exponent $\theta$ increases, while it is independent from the range of the hopping. 
	%\\
	The same scaling holds for the nonadiabatic work losses, i.e. the difference between the adiabatic work $W_\infty$ extracted in an infinitely slow cycle, and the work $W$ extracted in the more realistic finite-time case. Indeed, this difference can be expressed as\,\cite{Quan07PRE76}
	\begin{align}
		W_\infty-W = \sum_{k>0} \left[2\omega_{1,k}f_{2,k}+2\omega_{2,k}f_{1,k}\right](1-P_k)\label{eq: W_infty-W}
	\end{align} 
	which, in the optimal regime for the heat engine ($T_2 \ll \bar{\omega} \ll T_1$) becomes
	\begin{align}
		W_\infty-W \simeq 2\sum_{k>0}\omega_{1,k}(1-P_k).\label{eq: W_infty-W optimal}
	\end{align}
	As $\omega_{1,k}$ remains finite for $k \rightarrow 0$, the above expression has the same scaling of Eq.~\eqref{eq: excitation density}, as $\delta\to 0$,  
	\begin{align}
		W_\infty-W \propto  \delta^{\theta}  \sim n_{\rm ex},\label{eq: work_universal_scaling}
	\end{align}
	This result suggests that the actual leading physical mechanism behind the energy losses in a finite time cycle is the defect proliferation induced by the driving.
	\\
	Let us notice that, from the definition of the exponent $\theta$ in Eq.~\eqref{eq: kibble_zurek_exponent}, as the system is sufficiently long-range ($\alpha<2$), then $\theta_{LR} = 1/(2\alpha-2)>\theta_{SR} = 1/2$. This simple observation tells us that, at least in the limit of a slow cycle $\delta\to 0$, dynamical excitations are suppressed when long-range interactions are present. This additional source of long-range advantage mitigates one of the main limitations of quantum thermal devices, namely the trade-off between power and efficiency. Moreover, let us stress the fact that the result in Eq.~\eqref{eq: work_universal_scaling} only depends on universal quantities, suggesting that the reported long-range advantage is present in generic long-range interacting systems independently of the microscopic details of the model. Figure \ref{fig: EnergyLoss} shows the nonadiabatic work loss ratio $1-W/W_{\infty}$ as a function of $\delta$, for different values of $\alpha$. We notice that, excellent agreement is found between the exact numerical data (scatter plots) and the approximated result~\eqref{eq: W_infty-W optimal} we used to extract the universal scaling in Eq.~\eqref{eq: work_universal_scaling}. Moreover, as predicted by our analytical results, the work losses are widely reduced when $\alpha<2$.

	Finally, a similar reasoning applies to the refrigerator [R] in its most realistic temperature setting $T_2\lesssim T_1\ll \bar{\omega}$. As discussed in Sec. \ref{sec: Refrigerator operation mode}, in this case, the relevant quantity to be optimized is the heat extracted from the cold reservoir $Q_2$. Let us then consider the difference between the adiabatic cooling capability $Q_{2,\infty}$ and the heat extracted in a finite time cycle. In the range of temperatures $T_2\lesssim T_1\ll \bar{\omega}$ this reads
	%\begin{align}
	%	Q_{2,\infty}-Q_2 = 2\sum_{k>0}\omega_{2,k}f_{1,k}(1-P_k).
	%\end{align}
	%Then in the low temperatures limit $T_1\ll\mathcal{E}$ we find
	\begin{align}
		Q_{2,\infty}-Q_2 \simeq 2\sum_{k>0}\omega_{2,k}(1-P_k).
	\end{align}
	In order to determine the scaling of this quantity for a slow cycle ($\delta\to 0$) we have to distinguish the case in which $h_2$ is critical or not. While in the latter case we find the same result of Eq.~\eqref{eq: work_universal_scaling}, for $h_2 = 1$ (i.e. the ferromagnetic critical point) the dynamical scaling is affected by the presence of soft modes in $\omega_{2,k}$ as well, namely $\omega_{2,k} \sim k^{\min(\alpha_1,\alpha_2,2) - 1}$ (see Appendix \ref{app: Taylor expansion}). We find then the two different scaling behaviors
	\begin{align}
		Q_{2,\infty}-Q_2 \propto\begin{cases}
			\delta^\theta &h_2 \neq 1\\
			\delta^{\theta \min(\alpha_1,\alpha_2,2)} &h_2 = 1
		\end{cases}.\label{eq: Q2_universal_scaling}
	\end{align}
	Let us notice that, away from criticality, the performances of the thermal machine are enhanced for all $\alpha_2 < 2$, while for $h_2 = 1$ we have to require the additional constraint $\alpha_1 > 2 \alpha_2 - 2$ (which is trivially satisfied in the limits $\alpha_1 = \infty$, $\alpha_1 = \alpha_2 <2$). We conclude that we can have long-range advantage for the cooling capability of a finite time cycle as well.
	\section{Conclusion}
	In this work, we have investigated the performance of a quantum thermal machine consisting of a chain of fermions with power-law decaying interactions, which undergoes a quantum Otto cycle. We exactly computed the energy exchanged during the cycle, which in turn allowed us to provide a fully-fledged characterization of the device. We determined the regions of the parameter space corresponding to the most useful operation modes for quantum technological applications, i.e., the heat-engine $[E]$ and the refrigerator $[R]$ modes. Focusing on these two operation modes we then investigated the role of long-range interactions while optimizing the device performances, detecting several sources of long-range advantage with respect to the corresponding nearest-neighbor case. 
	
	Most remarkably, those results in high thermodynamic efficiency even when operating at finite power. Indeed the long-range character of the interactions is known to hinder the proliferation of defects as the critical point is crossed\,\cite{DefenuPRB2019}. We show that this mechanism is able to mitigate the energy losses, resulting in an advantage with respect to the short-range counterpart of the device. In particular, we were able to link such losses to the universal scaling of the defects within the Kibble-Zurek picture\,\cite{ZurekPRl2005}, suggesting this mechanism to be indeed very general. More precisely, as shown in Sec.\,\ref{sec: Finite time cycle}, the nonadiabatic energy losses in the work output~\eqref{eq: work_universal_scaling}, for the $[E]$ operation, and in the cooling capability~\eqref{eq: Q2_universal_scaling}, for the $[R]$ operation, are proportional to the density of dynamically generated defects during the evolution. This, in turn, is suppressed by the presence of sufficiently long-range interactions ($\alpha<2$). Indeed, the presence of long-range couplings results in a faster power law decay of the density of defects as the speed of the driving goes to zero, see Eq.~\eqref{eq: kibble_zurek_exponent}. The universality of the resulting long-range advantage makes systems with long-range interactions promising platforms for the implementation of finite-time many-body quantum thermal cycles with improved performances.

	Aside from nonadiabatic losses, long-range couplings in the fermionic chain were found to boost the performance of quantum thermodynamic machines even in the adiabatic regime. In particular, the optimal regime in the $[E]$ operation mode can be obtained by maximizing both the work output and the engine efficiency simultaneously. For a large temperature gradient between the thermal reservoirs, the optimal performance is achieved as the exponent of the power law decaying interaction is decreased. This effect is clearly demonstrated by the plots in Fig. \ref{fig: W_eta_a}, where the work output and the engine efficiency are shown to have a larger optimal value as the exponent $\alpha$ of the power law decaying interaction is decreased. Indeed, long-range interactions generate cusps in the low-energy spectrum of the fermionic quasi-particle that increase the work output of the engine, as follows already from the Einstein approximation, in which only the average level spacing is considered.
	
	At variance, in the refrigerator operation mode $[R]$, a realistic implementation requires the two baths to be close in temperature ($T_{2}\lesssim T_{1}$) and both deep in the quantum regime ($T_{1},T_{2}\ll \bar{\omega}$). This in turn results in a peak of the cooling capability corresponding to the quantum critical points of the model for any interaction range. The performance of the refrigerator close to the quantum critical points improves even further at small values of $\alpha$ as clearly indicated by Fig.\,\ref{fig: Q2_a10},  which yields a clear proof of the long-range advantage.
	
	The thermodynamic advantage demonstrated in the present analysis shall become even more prominent in the so-called strong long-range regime ($\alpha_{1,2} < 1$), in which the system loses additivity. However, the thermalization of the system on an accessible time scale is not guaranteed in this regime, as the fermionic dispersion relation becomes gapped\,\cite{DefenuPNAS2021, GiachettiArXiv2021}, thus allowing for the presence of long-lived prethermal and quasi-stationary phases\,\cite{AntoniPRE1995}. Further investigation is thus needed in order to understand this regime. However, we still may expect to find some source of thermodynamic advantage similar to the one obtained in Ref. \cite{YungerHalpernPRB2019} where the lack of thermalization is due to the presence of many-body localization.
	
	Finally, our findings could be experimentally checked in nowadays available trapped-ions platforms, which are currently used to realise long-range interacting quantum systems. The presence of a long-range advantage could thus lead to new quantum technological developments, with direct application to the cooling of quantum computers. 
	
	\subsection*{Acknowledgements}
	MC acknowledges useful discussions with G. Piccitto and D. Rossini
	on a problem that is closely related to the present work.  We acknowledge support by the Deutsche Forschungsgemeinschaft (DFG, German Research Foundation)   under Germany’s Excellence Strategy 
	EXC2181/1-390900948 (the Heidelberg STRUCTURES Excellence Cluster). This work is part of the MIUR-PRIN2017 project Coarse-grained description for nonequilibrium systems and transport phenomena (CO-NEST) No. 201798CZL.

	\newpage
	\appendix
		\section{Nonadiabatic energy exchanges}
		In this Appendix we provide the details of the derivation of the explicit expressions for the engine energy exchanges in the generic nonadiabatic case, reported in Eqs.~\eqref{eq: nonadiabatic exchanges} of the main text. Let us start from the definition of the energy exchanges with the two thermal reservoirs involved in the cycle
		\begin{align}
			Q_1 &= E_1-E_4 = \mathrm{Tr}[\rho_1H_1]-\mathrm{Tr}[\rho_4H_1],\\
			Q_2 &= E_3-E_2 = \mathrm{Tr}[\rho_3 H_2]-\mathrm{Tr}[\rho_2H_2].
		\end{align}
		Since $E_1$ and $E_3$ are thermal expectations they can be readily expressed as 
		\begin{align}
			E_1 = -\sum_{k>0}\omega_{k,1}\tanh\left(\frac{\beta_1\omega_{k,1}}{2}\right),\quad E_3 = -\sum_{k>0}\omega_{k,2}\tanh\left(\frac{\beta_2\omega_{k,2}}{2}\right).\label{eq:E1E3}
		\end{align}
		In order to compute $E_2$ and $E_4$, we notice that, as shown in Section \ref{sec: Finite time cycle}, the unitary evolution generated by $H(h(t))$ is such that it only mixes the states $|0_{k},0_{-k}\rangle$ and $|1_{k},1_{-k}\rangle$, for each value of $k$. As a consequence, the dynamics of the Kitaev chain can be exactly described by the unitary dynamics of $N$ independent two level systems, i.e., the unitary evolution operator can be decomposed as
		\begin{align}
			U = \bigotimes_k U_k,\quad U_k = \mathrm{T}\exp\left[-i\int_0^\tau \mathcal{H}_k(t)dt\right],
		\end{align}
		where the unitary operator associated to each two level system is generated by the time dependent Hamiltonian
		\begin{align}
			\mathcal{H}_k(t) = \frac{\omega_k(t)}{2}(|\phi^{+}_k(t)\rangle\langle \phi^{+}_k(t)|-|\phi^{-}_k(t)\rangle\langle \phi^{-}_k(t)|),
		\end{align}
		where $|\phi^{\pm}_k\rangle$ are the instantaneous eigenstates of the Hamiltonian defined in Eq.\eqref{eq: istantanesous eigenstates}.
		The energies $E_2$ and $E_4$ at the end of the two unitary strokes of the cycle can then be written as
		\begin{align}
			E_2 &= \sum_{k}\mathcal{E}_{k,2}, \quad \mathcal{E}_{k,2} = \mathrm{Tr}[U_ke^{-\beta_1\mathcal{H}_{k,1}}U_k\mathcal{H}_{k,2}]/Z_{k,1},
			\\
			E_4 &= \sum_{k}\mathcal{E}_{k,4}, \quad \mathcal{E}_{k,2} = \mathrm{Tr}[\tilde{U}_ke^{-\beta_2\mathcal{H}_{k,2}}\tilde{U}_k\mathcal{H}_{k,1}]/Z_{k,2},
		\end{align}
		where $\mathcal{E}_{k,i}$ is the energy associated to the $k$th mode, which can be computed by using the results for the single qubit Otto cycle of Ref.\cite{SolfanelliPRB2020}, which we briefly summarize here for the sake of completeness. In particular, for the energies $\mathcal{E}_{k,i}$ we find
		\begin{align}
			\mathcal{E}_{k,2} &= \sum_{a,b = \pm}\frac{e^{-\beta_1\epsilon^{(a)}_{k,1}}\epsilon_{k,2}^{(b)}|\langle\phi_{k,2}^{(b)}|U_k|\phi_{k,1}^{(a)}\rangle|^2}{2\cosh(\beta_1\omega_{1,k}/2)},
			\\
			\mathcal{E}_{k,4} &= \sum_{a,b = \pm}\frac{e^{-\beta_2\epsilon^{(a)}_{k,2}}\epsilon_{k,1}^{(b)}|\langle\phi_{k,1}^{(b)}|\tilde{U}_k|\phi_{k,2}^{(a)}\rangle|^2}{2\cosh(\beta_1\omega_{2,k}/2)},
		\end{align}
		where $\epsilon_{k,i}^{(\pm)} = \pm\omega_{i,k}/2$, for $i = 1,2$. Note that the $2\times 2$ square matrix $P_{a,b}^k = |\langle\phi_{k,2}^{(b)}|U_k|\phi_{k,2}^{(a)}\rangle|^2$ is doubly stochastic, namely, $0\leq P_{a,b}^k\leq 1$, $\sum_a P_{a,b}^k=
		\sum_b P_{a,b}^k = 1$. This immediately implies that one of its elements is sufficient to determine all of them, and that the matrix is symmetric: if $P_{++}^k = P_k$, then $P_{+-} = P_{-+} = 1-P_k$, and $P_{--} = P_k$. Moreover,as shown in Ref. \cite{SolfanelliPRB2020}, thanks to the fact that $\mathcal{H}_{k}(t)$ is invariant under the antiunitary complex conjugation operator we have that $\tilde{P}_k = P_k$. Hence, $\mathcal{E}_{k,2}$ and $\mathcal{E}_{k,4}$ read
		\begin{align}
			\mathcal{E}_{k,2} = \frac{\omega_{2,k}}{2}\tanh\left(\frac{\beta_1\omega_{1,k}}{2}\right)(1-2P_k),
			\quad
			\mathcal{E}_{k,4} = \frac{\omega_{1,k}}{2}\tanh\left(\frac{\beta_2\omega_{2,k}}{2}\right)(1-2P_k).
		\end{align}
		Then, summing over the $k$-modes and using the symmetry for $k\to -k$ of the Hamiltonian we obtain
		\begin{align}
			E_2 = \sum_{k>0}\omega_{2,k}\tanh\left(\frac{\beta_1\omega_{1,k}}{2}\right)(1-2P_k),\quad
			E_4 = \sum_{k>0}\omega_{1,k}\tanh\left(\frac{\beta_2\omega_{2,k}}{2}\right)(1-2P_k).\label{eq:E2E4}
		\end{align}
		Finally, inserting the expression for $E_1$ and $E_3$ in Eq.~\eqref{eq:E1E3} and those for $E_2$ and $E_4$ in Eq.~\eqref{eq:E2E4} into the definition of the heat exchanges we find
		\begin{align}
			Q_1 &= -\sum_{k>0}\omega_{1,k}\left[\tanh\left(\frac{\beta_1\omega_{k,1}}{2}\right)+\tanh\left(\frac{\beta_2\omega_{2,k}}{2}\right)(1-2P_k)\right],
			\\
			Q_2 &= -\sum_{k>0}\omega_{2,k}\left[\tanh\left(\frac{\beta_2\omega_{k,2}}{2}\right)+\tanh\left(\frac{\beta_1\omega_{1,k}}{2}\right)(1-2P_k)\right],
		\end{align}
		which reduce to the expressions in Eqs.~\eqref{eq: Q1 nonadiabatic} and \eqref{eq: Q2 nonadiabatic} once the coefficients $f_{k,i} = \tan(\beta_i\omega_{i,k}/2)$ are introduced.
	\section{Taylor expansion of the spectrum around the critical modes}\label{app: Taylor expansion}
	In this Appendix we compute the Taylor expansion of the quasiparticle spectrum~\eqref{eq: spectrum} at lowest order in $|k-k_c|)$, where $k_c = 0$ at the critical point $h = 1$, while $k_c = \pi$ at $h = -1+2^{1-\alpha_1}$. Lets start from the $h = 1$ critical point, for long-range couplings with $1<\alpha_1<3$ and $1<\alpha_2<2$, at lowest order in $k\simeq 0$  we find\,\cite{DefenuPRB2019}
	\begin{align}
		&t_k = 1+A(\alpha_1)k^{\alpha_1-1}+O(k^2),
		\\
		&\Delta_k = B(\alpha_2)k^{\alpha_2-1}+O(k),
	\end{align}
	where we have introduced $A(\alpha) = \sin(\alpha\pi/2)\Gamma(1-\alpha)/\zeta(\alpha)$ and $B(\alpha) = \cos(\alpha\pi/2)\Gamma(1-\alpha)/\zeta(\alpha)$. While in the short-range case $\alpha_1 = \alpha_2 = \infty$ we simply have
	\begin{align}
		&t_k = \cos(k) = 1-\frac{k^2}{2}+O(k^4),
		\\
		&\Delta_k = \sin(k) = k +O(k^2).
	\end{align}
	Inserting these expansions in Eq. \ref{eq: spectrum} we obtain  
	\begin{align}
		\omega_k = \begin{cases}
			|h-1| + O(k^{\alpha-1})	&h\neq 1
			\\
			C(\alpha) k^{\alpha-1}+O(k^{2\alpha-2}) &h= 1,
		\end{cases}\label{eq: dispersion relation 0}
	\end{align}  
	where $\alpha = \min(\alpha_1,\alpha_2)$ and
	\begin{align}
		C(\alpha) = \begin{cases}
			A(\alpha_1) &\mathrm{if}\quad\alpha_1<\alpha_2\\
			\sqrt{A^2(\alpha)+B^2(\alpha)} &\mathrm{if}\quad\alpha_1=\alpha_2=\alpha \\
			B(\alpha_2) &\mathrm{if}\quad\alpha_1>\alpha_2\\
		\end{cases}.\label{eq: prefactor 0}
	\end{align}
	%
	%If the hopping is short-range $\alpha_1 = \infty$ and the pairing is long-range $1<\alpha_2 = \alpha<2$, we find
	%
	%\begin{align}
	%	\omega_k = \begin{cases}
	%		|h-1| + O(k^{2\alpha-2})	&h\neq 1
	%		\\
	%		\cos(\frac{\alpha\pi}{2})\frac{\Gamma(1-\alpha)}{\zeta(\alpha)}k^{\alpha-1}+O(k^2) &h= 1.
	%	\end{cases}
	%\end{align}
	%
	On the other hand in the nearest neighbor case $\alpha_1 = \alpha_2 = \infty$, we obtain 
	\begin{align}
		\omega_k = \begin{cases}
			|h-1| + O(k^{2})	&h\neq 1
			\\
			k+O(k^2) &h= 1.
		\end{cases}
	\end{align}
	The Taylor expansions around $k = \pi$ are obtained from the results at $k = 0$, using the following relation 
	\begin{align}
		\mathrm{Li}_\alpha(ze^{i\pi}) = 2^{1-\alpha}\mathrm{Li}_\alpha(z^2)-\mathrm{Li}_\alpha(z).
	\end{align}
	Then applying this property of the polylogarithm to the definition of $t_k$ and $\Delta_k$ in Eqs.~\eqref{eq: hopping},~\eqref{eq: pairing}, we find
	\begin{align}
		&t_k = 2^{1-\alpha_1}t_{2(k-\pi)}-t_{k-\pi}\label{eq: tkpi}
		\\
		&\Delta_k = 2^{1-\alpha_2}\Delta_{2(k-\pi)}-\Delta_{k-\pi}.\label{eq: Dkpi}
	\end{align}
	The Taylor expansion of $t_k$ and $\Delta_k$ around $k=\pi$ follows by applying the expansion around $k'= 0$ to $t_{k'}$ and $\Delta_{k'}$ with $k' = 2(k-\pi)$ and $k'=k-\pi$, respectively, then inserting the results in Eqs.~\eqref{eq: tkpi},~\eqref{eq: Dkpi}. Then for $k\simeq \pi$, $1<\alpha_1<3$ and $1<\alpha_2<2$, we find 
	\begin{align}
		t_k &= -1+2^{1-\alpha_1}-D(\alpha_1)(\pi-k)^2\notag\\&+O((\pi-k)^3),
		\\
		\Delta_k &= E(\alpha_2)(\pi-k)\notag\\&+O((\pi-k)^3),
	\end{align}
	where $D(\alpha_1) = (2^{3-\alpha_1}-1)\zeta(\alpha_1-2)/2\zeta(\alpha_1)$ and $E(\alpha) =(1-2^{2-\alpha_2})\zeta(\alpha_2-1)/\zeta(\alpha_2))$. While in the short-range case we have 
	\begin{align}
		&t_k =-1+\frac{(\pi-k)^2}{2}+O((\pi-k)^4),
		\\
		&\Delta_k =\pi-k +O((\pi-k)^3).
	\end{align}
	Inserting these results in Eq.~\eqref{eq: spectrum} we obtain, for $\alpha_1=\alpha_2=\alpha$, and $1<\alpha<2$
	\begin{align}
		\omega_k = \begin{cases}
			|h+1-2^{1-\alpha_1}| + O((\pi-k)^{2})	&h\neq h_c(\alpha)
			\\
			F(\alpha)(\pi-k)+O((\pi-k)^3) &h= h_c(\alpha),
		\end{cases}\label{eq: dispersion relation pi}
	\end{align}
	with $\alpha = \min(\alpha_1,\alpha_2)$ and
	\begin{align}
		F(\alpha) = \begin{cases}
			D(\alpha_1) &\mathrm{if}\quad\alpha_1<\alpha_2\\
			\sqrt{D^2(\alpha)+E^2(\alpha)} &\mathrm{if}\quad\alpha_1=\alpha_2=\alpha \\
			E(\alpha_2) &\mathrm{if}\quad\alpha_1>\alpha_2\\
		\end{cases}.\label{eq: prefactor pi}
	\end{align}
	
	%If $\alpha_1 = \infty, 1<\alpha_2 = \alpha<2$, then
	%
	%\begin{align}
	%	\omega_k = \begin{cases}
	%		|h+1| + O((\pi-k)^{2})	&h\neq -1
	%		\\
	%		\frac{\zeta(\alpha-1)}{\zeta(\alpha)}(2^{2-\alpha}-1)(\pi-k)+O((\pi-k)^3) &h=-1.
	%	\end{cases}
	%\end{align}
	%
	Finally, in the short-range case $\alpha_1 = \alpha_2 = \infty$ we find
	\begin{align}
		\omega_k = \begin{cases}
			|h+1| + O((\pi-k)^{2})	&h\neq -1
			\\
			(\pi-k)+O((\pi-k)^3) &h=-1.
		\end{cases}
	\end{align}
		\section{Work output in the infinite temperature gradient limit}\label{app: Work output in the infinite temperature gradient limit}
		In this Appendix we provide the derivation of Eq.~\eqref{eq: W0} for the work output in the limit of infinite hot temperature $T_1\gg \bar{\omega}(h)$ and infinitely small cold temperature $T_2\ll \bar{\omega}(h)$, with respect to the typical energy scale of the system which we identify with the average spectrum~\eqref{eq: average spacing}. Let us start from the expression of the energy exchanges of the adiabatic cycle in Eqs.~\eqref{eq: Q1 adiabatic},~\eqref{eq: Q2 adiabatic} , and ~\eqref{eq: W adiabatic}, then we have to expand the coefficients $f_{k,1}$ and $f_{k,2}$ for $\omega_{1,k}\beta_1 = \omega_{1,k}/T_1\to 0$ and $\omega_{2,k}\beta_2 = \omega_{2,k}/T_2 \to \infty$, respectively. In particular, at leading order we find 
		\begin{align}
			f_{k,1} = \mathcal{O}(\beta_1\omega_{1,k}),\quad f_{k,2} = 1-\mathcal{O}(e^{-\beta_2\omega_{2,k}}).
		\end{align}
		Then, inserting this expressions into Eqs.~\eqref{eq: Q1 adiabatic} and~\eqref{eq: Q2 adiabatic} we obtain
		\begin{align}
			Q_1 &= \sum_{k>0}\omega_{1,k}(1+\mathcal{O}(\beta_1\omega_{1,k})+\mathcal{O}(e^{-\beta_2\omega_{2,k}}))
			\\
			Q_2 &= -\sum_{k>0}\omega_{2,k}(1+\mathcal{O}(\beta_1\omega_{1,k})+\mathcal{O}(e^{-\beta_2\omega_{2,k}}))
		\end{align}
		Finally, using the first principle of thermodynamics we find the work output
		\begin{align}
			W = \sum_{k>0}(\omega_{1,k}-\omega_{2,k})(1+\mathcal{O}(\beta_1\omega_{1,k})+\mathcal{O}(e^{-\beta_2\omega_{2,k}}))\approx W_0,
		\end{align}
		with
		\begin{align}
			W_0 = \sum_{k>0}(\omega_{1,k}-\omega_{2,k}) = \frac{\bar{\omega}_1-\bar{\omega}_2}{2}.
		\end{align}
		The effect of higher order corrections, leading to the advantage-disadvantage transition outside from the optimal region, is considered in Appendix \ref{app: Advantage-disadvantage transition}.
	\section{Cooling capability in the low temperature limit}\label{app: Cooling capability}
	In this Appendix we provide the detailed derivation of Eq.~\eqref{eq: Q2 power law} in the main text for the low-temperature limit of the cooling capability when the device works as a refrigerator. We start from expression~\eqref{eq: Q2 adiabatic}  for the heat extracted from the cold reservoir in an adiabatic cycle. Then the leading contribution as $T_2\lesssim T_1\ll\bar{\omega}$ is given by
	\begin{align}
		Q_2 \simeq \sum_{k>0}\omega_{2,k}e^{-\beta_2\omega_{2,k}}.
	\end{align}
	For large system size $N\gg 1$ we can perform a continuum limit passing from a sum over the $k$-modes to an integral 
	\begin{align}
		\frac{Q_2}{N} \simeq \int_0^\pi\frac{dk}{\pi}\omega_{2,k}e^{-\beta_2\omega_{2,k}}.
	\end{align}
	Since $T_2\to 0$ ($\beta_2\to \infty$), the leading contribution to the integral comes from the modes at which $\omega_{2,k}$ is minimum. Then if $h_2\neq 1,-1+2^{1-\alpha_1}$, the spectrum is gapped and $\min_k[\omega_{2,k}]>0$. Accordingly, at leading order, we have the exponential behavior
	\begin{align}
		\frac{Q_2}{N} \simeq \min_k[\omega_{2,k}]e^{-\beta_2\min_k[\omega_{2,k}]}.
	\end{align}
	On the other hand at the quantum critical points,  $\min_k\omega_{2,k} =0$, leading to a power law decay. In particular at the ferromagnetic critical point $h_2 = 1$, using the dispersion relation in Eq.~\eqref{eq: dispersion relation 0}, we obtain
	\begin{align}
		\frac{Q_2}{N}\simeq \int_0^\pi\frac{dk}{\pi}C(\alpha)k^{\alpha-1}e^{-\beta_2 C(\alpha)k^{\alpha-1}},
	\end{align}
	where $\alpha = \min(\alpha_1,\alpha_2)$ and the $C(\alpha)$ is given by Eq.~\eqref{eq: prefactor 0}. Finally performing the change of variables $y = \beta_2C(\alpha)k^{\alpha-1}$ we find
	\begin{align}
		\frac{Q_2}{N}\simeq T_2^{\frac{\alpha}{\alpha-1}}C(\alpha)^{\frac{1}{\alpha-1}}\Gamma(\alpha/(\alpha-1))/\pi,
	\end{align}
	where we have introduced the gamma function $\Gamma(x) = \int_0^\infty dyy^{x-1}e^{-y}$. Analogously, at the antiferromagnetic critical point, using the dispersion relation in Eq.~\eqref{eq: dispersion relation pi}, we find
	\begin{align}
		\frac{Q_2}{N}\simeq T_2^2F(\alpha)/\pi.
	\end{align}
	Finally, introducing the dynamical critical exponent~\eqref{eq: dynamical exponent 1}, ~\eqref{eq: dynamical exponent 2}, we obtain the result of Eq.~\eqref{eq: Q2 power law}, the $\alpha$ dependent prefactor given by
	\begin{align}
		K(\alpha) = \begin{cases}
			C(\alpha)^{\frac{1}{\alpha-1}}\Gamma(\alpha/(\alpha-1))/\pi &h_2 = 1\\
			F(\alpha)/\pi &h_2 = -1+2^{1-\alpha}
		\end{cases}.
	\end{align}
	\begin{figure*}%[t!]
		\centering
		\includegraphics[width=\textwidth]{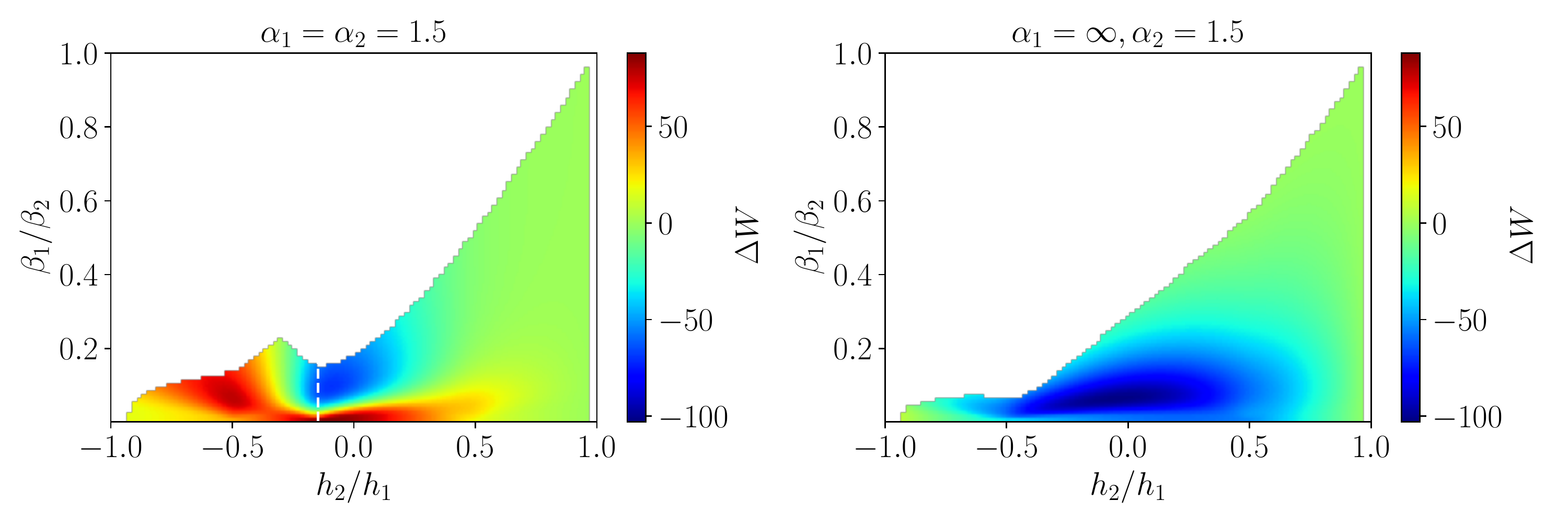}
		\centering
		\includegraphics[width=\textwidth]{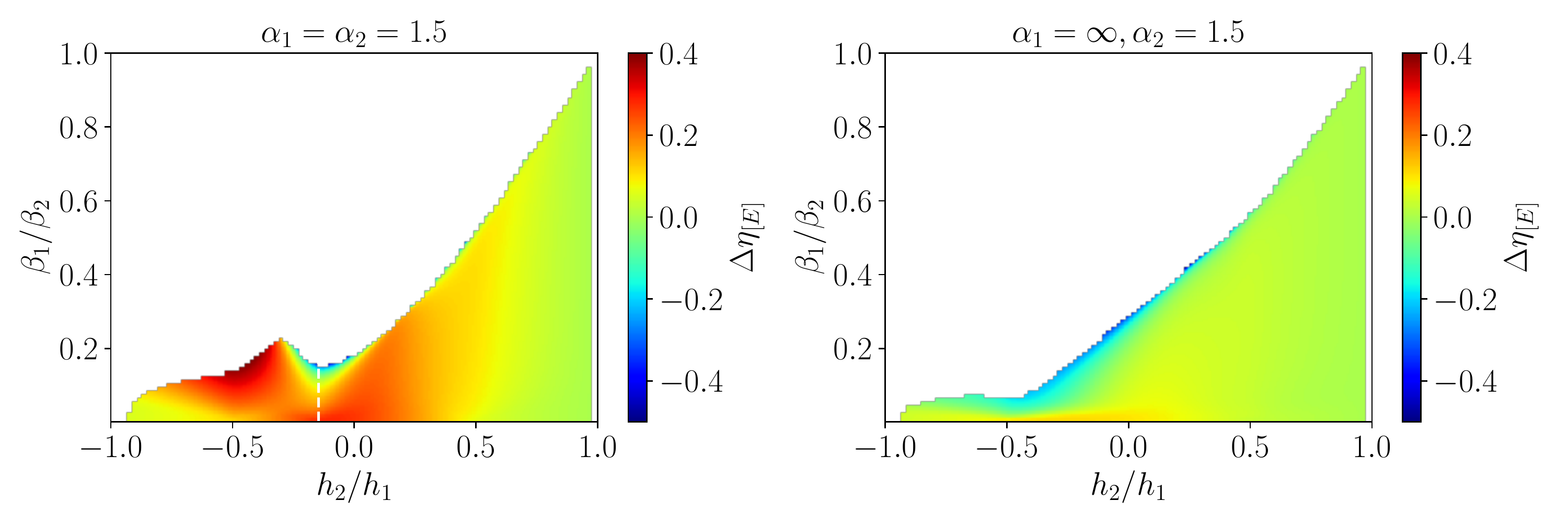}
		\caption{Advantage parameters $\Delta W$ and $\Delta_{\eta_{[E]}}$ plotted against $\beta_1/\beta_2$ and $h_2/h_1$ in the region of parameters corresponding to the heat-engine operation. The $\alpha_1=\alpha_2 =1.5$ case is compared with $\alpha_1 = \infty$, $\alpha_2 = 1.5$.}
		\label{fig: DW Deta}
	\end{figure*}
	\section{Characterization of the long-range advantage in the full parameter space}\label{app: Characterization of the long-range advantage}
	\begin{figure*}%[t!]
		\centering
		\includegraphics[width=\textwidth]{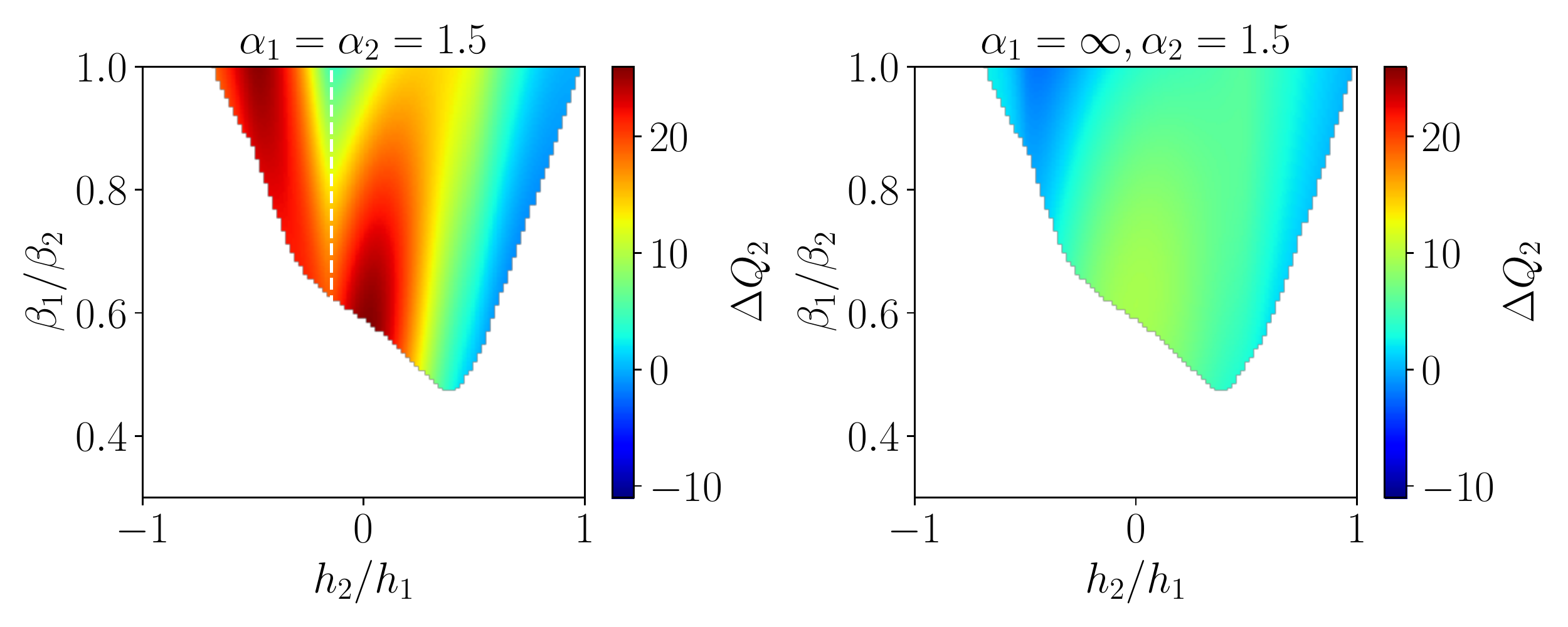}
		\caption{Advantage parameter $\Delta Q_2$ plotted against $\beta_1/\beta_2$ and $h_2/h_1$ in the region of parameters corresponding to the refrigerator $[R]$ operation. The $\alpha_1=\alpha_2 =1.5$ case is compared with $\alpha_1 = \infty$, $\alpha_2 = 1.5$.}
		\label{fig: DQ2}
	\end{figure*}
	In Sections \ref{sec: Heat Engine operation} and \ref{sec: Refrigerator operation mode} we have seen how long-range interactions lead to a substantial advantage in the optimal regimes of the heat engine $[E]$ and the refrigerator $[R]$ operations, respectively. However, a question may arise on whether such an advantage is preserved in an extended region of the parameters space, or if it is present only for those fine-tuned values. In this Appendix we address this question showing that the long-range advantage actually extends to a connected region of the parameter space before an advantage/disadvantage transition takes place. In particular, we can introduce the advantage parameters which for the $[E]$ operation are defined as 
	\begin{align}
		\Delta W &= W_{\mathrm{lr}}-W_{\mathrm{sr}},\label{eq: advantage parameter E}\\
		\Delta \eta &= \eta_{\mathrm{lr}}-\eta_{\mathrm{sr}},\label{eq: advantage parameter etaE}
	\end{align}
	where $W_{\mathrm{lr}(\mathrm{sr})}$ and $\eta_{\mathrm{lr}(\mathrm{sr})}$ are the engine work output and efficiency in the long-range(lr) and short-range(sr) case, respectively. Then, a long-range advantage is indicated by the conditions $\Delta W\geq 0$, $\Delta\eta\geq 0$, signaling larger work output and efficiency in the long-range case. Figure \ref{fig: DW Deta} shows the $\Delta W$~\eqref{eq: advantage parameter E} and $\Delta\eta$~\eqref{eq: advantage parameter etaE} parameters as functions of the relevant parameters $h_2/h_1$ and $\beta_1/\beta_2$, in the $[E]$ region (see fig. \ref{fig: Regimes}), and for different values of $\alpha_1$,$\alpha_2$. In particular, we notice that in the $\alpha_1 = \alpha_2$ case a connected advantage domain is present (red area in the plots of Fig. \ref{fig: DW Deta}) near to the optimal region with $\beta_1/\beta_2\ll 1$. Then, as the temperature ratio is increased, an advantage to disadvantage transition takes place, with the transition point identified by the condition $\Delta W = 0$. On the other hand for short-range pairing $\alpha_1 = \infty$, only minor advantages are present while the disadvantage region is much more extended with respect to the fully long-range case. This observation justifies our choice to focus mainly on the $\alpha_1 = \alpha_2$ case in this paper.
	
	Analogously for the $[R]$ operation, we can introduce an advantage parameter as the difference between the cooling capabilities $Q_2$, in the long-range and short-range cases
	\begin{align}
		\Delta Q_2 = Q_{2,\mathrm{lr}}-Q_{2,\mathrm{sr}}.
	\end{align}
	This quantity is plotted as a function of $\beta_1/\beta_2$ and $h_2/h_1$ and for different values of $\alpha_1$, $\alpha_2$, in Fig. \ref{fig: DQ2}. Also in this case we notice that an extended advantage region is present for long-range pairing and hopping $\alpha_1 = \alpha_2 = 1.5$ (see Fig. \ref{fig: DQ2}a)), while the advantage almost disappears in when the pairing is short-range $\alpha_1 = \infty$ (Fig. \ref{fig: DQ2}b)).
	\begin{figure}%[t!]
		\centering
		\includegraphics[width=0.45\textwidth]{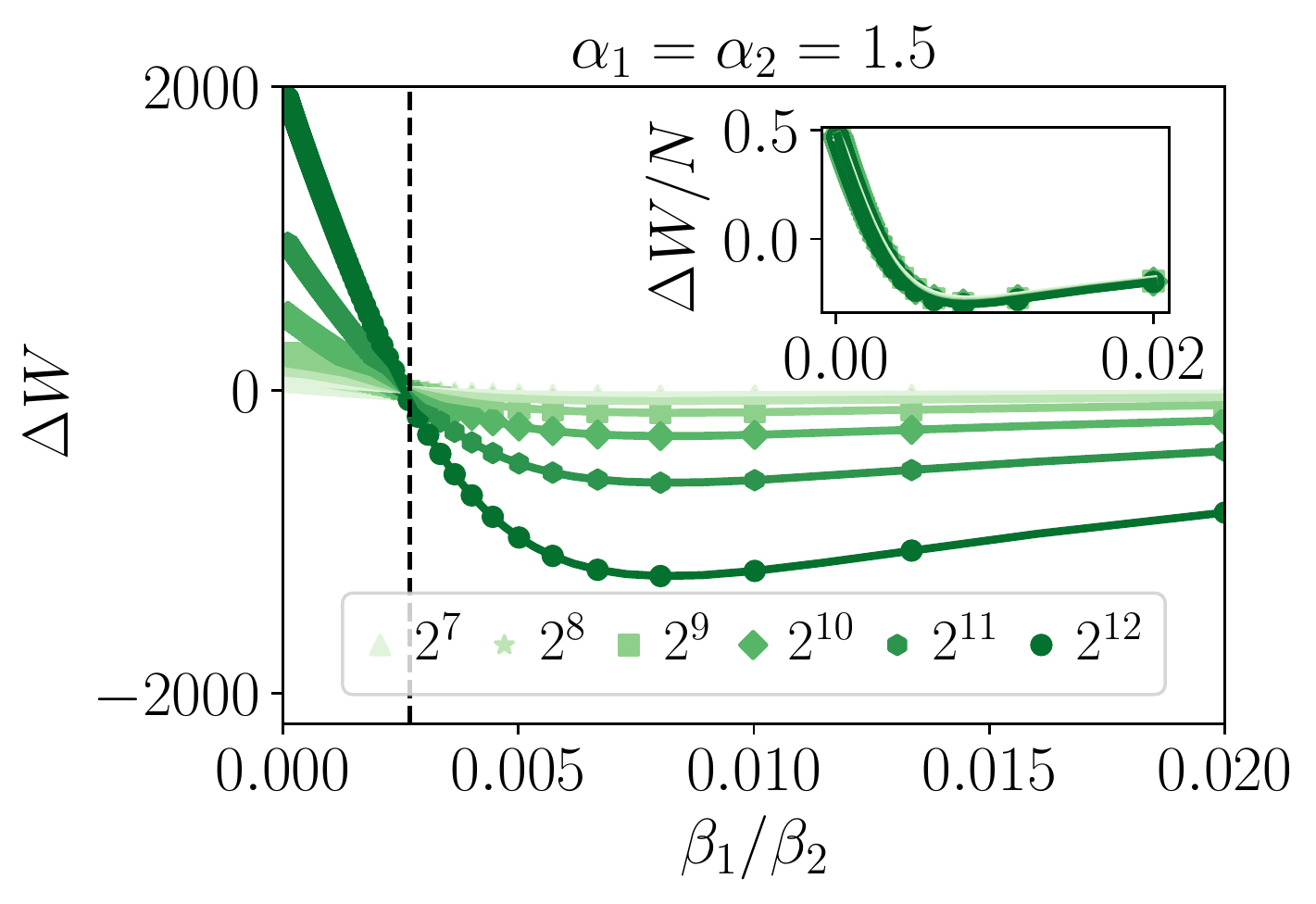}
		\caption{Scaling of $\Delta W$ for different values of the system size $N$, in the proximity of the advantage disadvantage transition point (black dashed line). In the inset $\Delta W/N$ is plotted, showing that the data corresponding to different system sizes collapse to the same curve.}
		\label{fig: DW_Scaling}
	\end{figure}
		\section{Hopping and Pairing contributions to the long-range advantage}
\begin{figure}%[t!]
			\centering
			\includegraphics[width=0.7\textwidth]{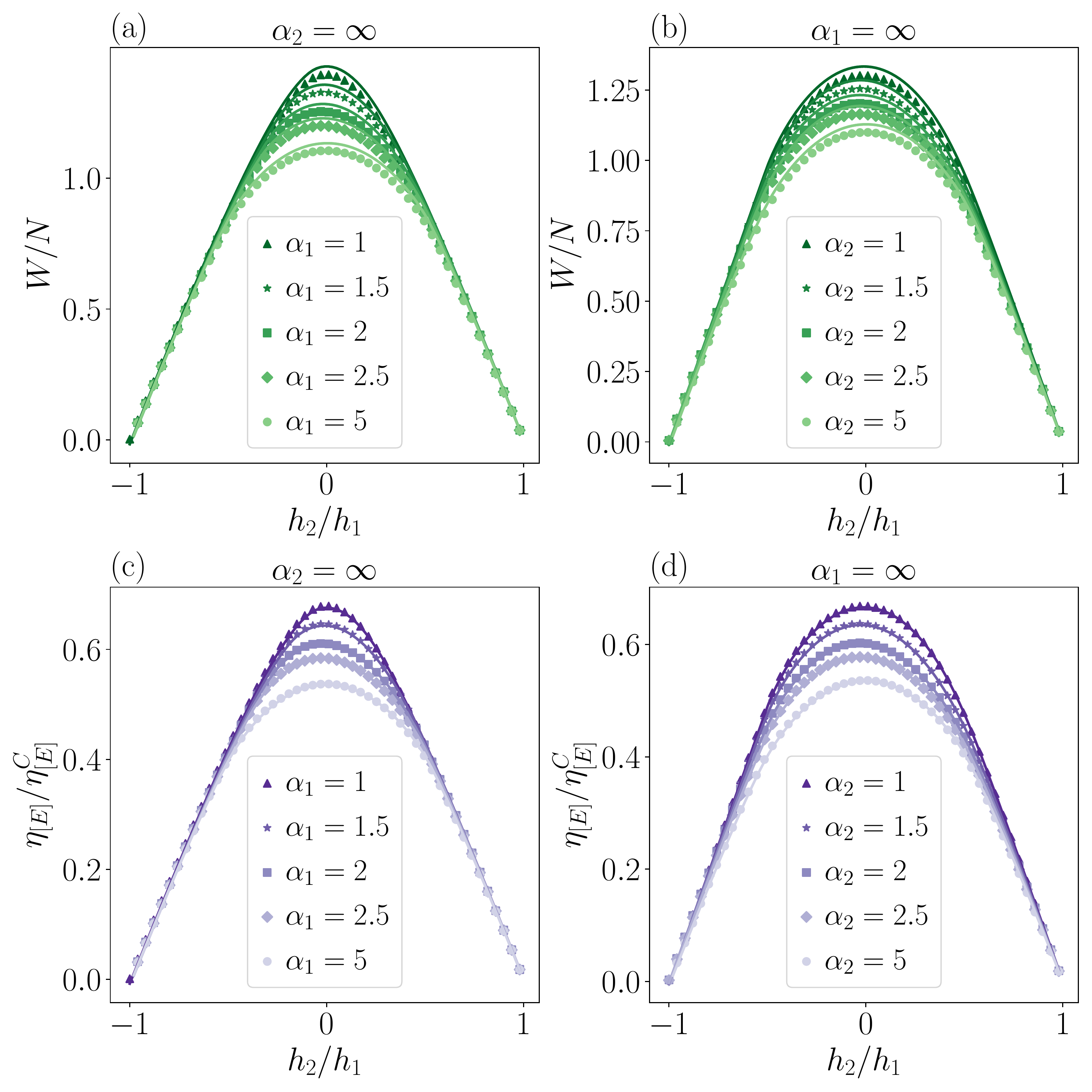}
			\caption{Same as Fig. \ref{fig: W_eta_a} but for $\alpha_2 = \infty$ and finite $\alpha_{1}$ in panels a) and c), and $\alpha_1 = \infty$ and finite $\alpha_{2}$ in panels b) and d).}
			\label{fig: W_eta_a12_h}
		\end{figure}

		In this, Appendix we extend the analysis of Sections \ref{sec: Heat Engine operation} to generic values of power law decay exponents $\alpha_{1}$ and $\alpha_{2}$. In particular, Fig. \ref{fig: W_eta_a12_h} shows the work output $W$ and the heat engine efficiency divided by the Carnot efficiency $\eta_{[E]}/\eta_{[E]}^C$, as a function of $h_2/h_1$ for the same parameter values of Fig. \ref{fig: W_eta_a} in this two opposite cases. We notice that, independently from the values of $\alpha_{1}$ and $\alpha_{2}$, both the work output and the engine efficiency have the same qualitative behavior, in particular they both present a maximum for $h_2/h_1\simeq 0$. Moreover, this optimal value is enhanced as the power law decay exponent of the long-range coupling ($\alpha_1$, $\alpha_{2}$ or both) is lowered, thus leading to a long-range advantage. Accordingly, the analysis of the three extremal cases with long-range hopping and short-range pairing (Fig.\ref{fig: W_eta_a12_h}a-c), long-range pairing and short-range hopping (Fig.\ref{fig: W_eta_a12_h}b-d), and equally long-range 
		pairing and hopping amplitudes (Fig. \ref{fig: W_eta_a}), suggests that a similar qualitative behavior may be found for any intermediate values of $\alpha_{1}$ and $\alpha_{2}$ leading to an advantage whenever at least one of the two couplings is sufficiently long-range. 
		\begin{figure}%[t!]
			\centering
			\includegraphics[width=0.8\textwidth]{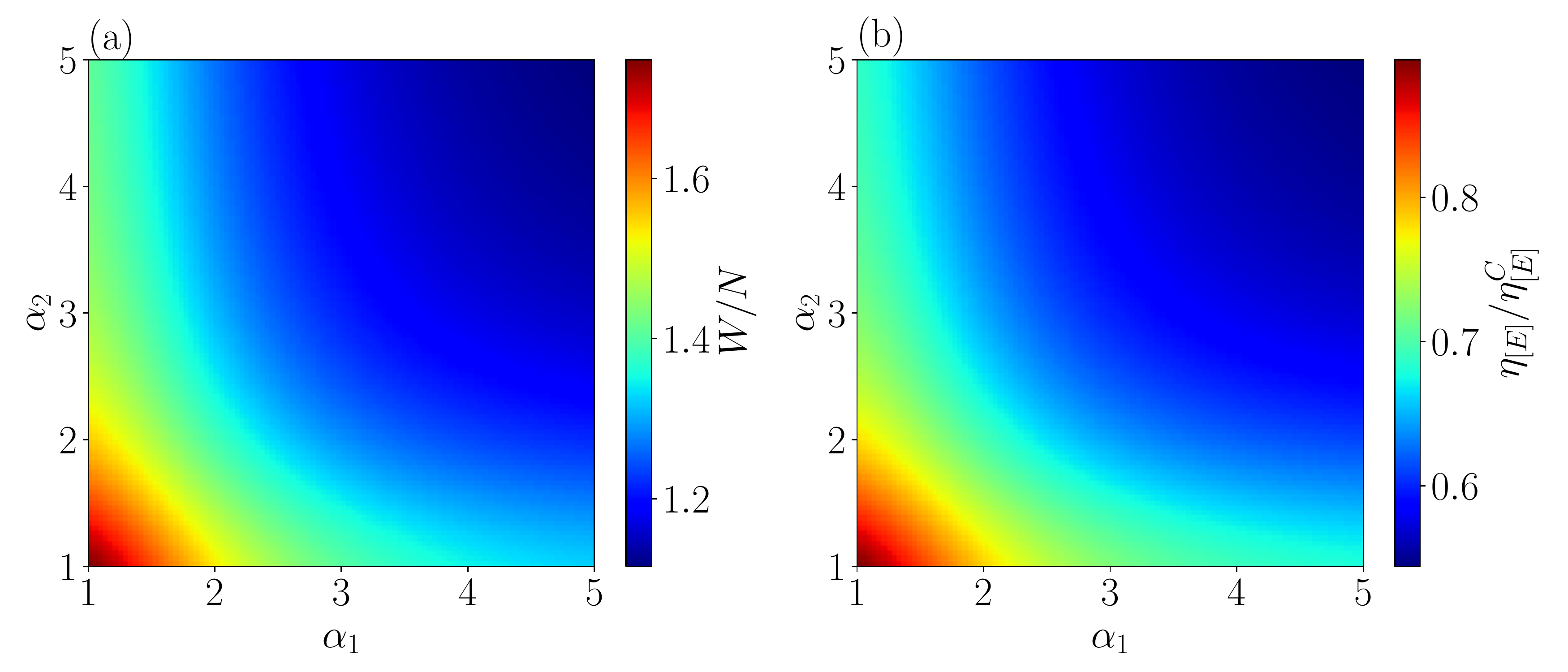}
			\caption{Work output (panel a) and engine efficiency (panel b) plotted as a function of $\alpha_{1}$ and $\alpha_{2}$, for fixed values of the cycle parameters in the optimal region: $h_1 = 2$, $h_2 = 0$, $T_1 = 100$, $T_2 = 0.01$, and for system size $N = 200$.}
			\label{fig: W_eta_a1_a2}
		\end{figure}
		On the other hand, the optimal values of $W$ and $\eta_{[E]}$, corresponding to their maxima at $h_2/h_1\approx 0$, are enhanced as the range of both couplings is increased, i.e., when $\alpha_{1},\alpha_{2}\to 1$. This is clearly shown in Fig. \ref{fig: W_eta_a1_a2}, where $W/N$ and $\eta_{[E]}/\eta_{[E]}^C$ are plotted as a function of $\alpha_{1}$ and $\alpha_{2}$, for $h_2/h_1 = 0$.  

\begin{figure}%[t!]
			\centering
			\includegraphics[width=\textwidth]{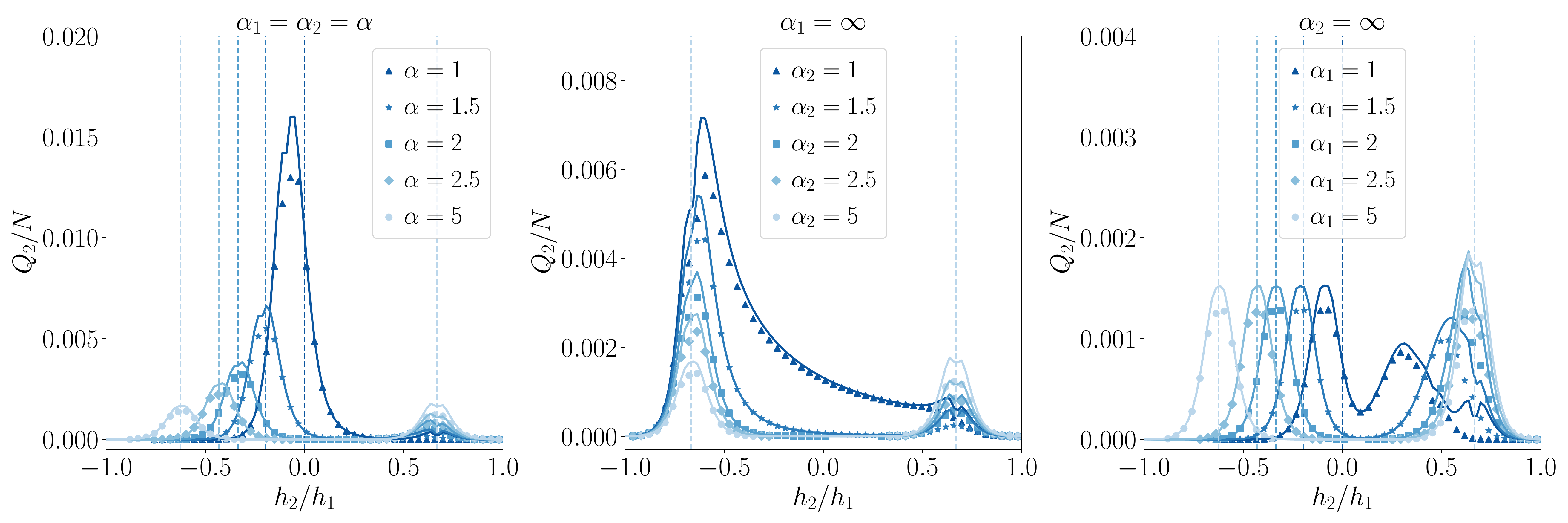}
			\caption{Same as Fig. \ref{fig: Q2_a10} but for different values of $\alpha_{1}$ and $\alpha_{2}$: finite $\alpha_{1} = \alpha_{2} = \alpha$ (panel a), $\alpha_1 = \infty$ and finite $\alpha_{2}$ (panel b), $\alpha_2 = \infty$ and finite $\alpha_{1}$ (panel c). Dashed vertical lines indicate the quantum critical points $h = 1$ and $h_2 = -1+2^{1-\alpha_{1}}$.}
			\label{fig: Q2_a12_h}
		\end{figure}

\begin{figure}%[t!]
			\centering
			\includegraphics[width=0.5\textwidth]{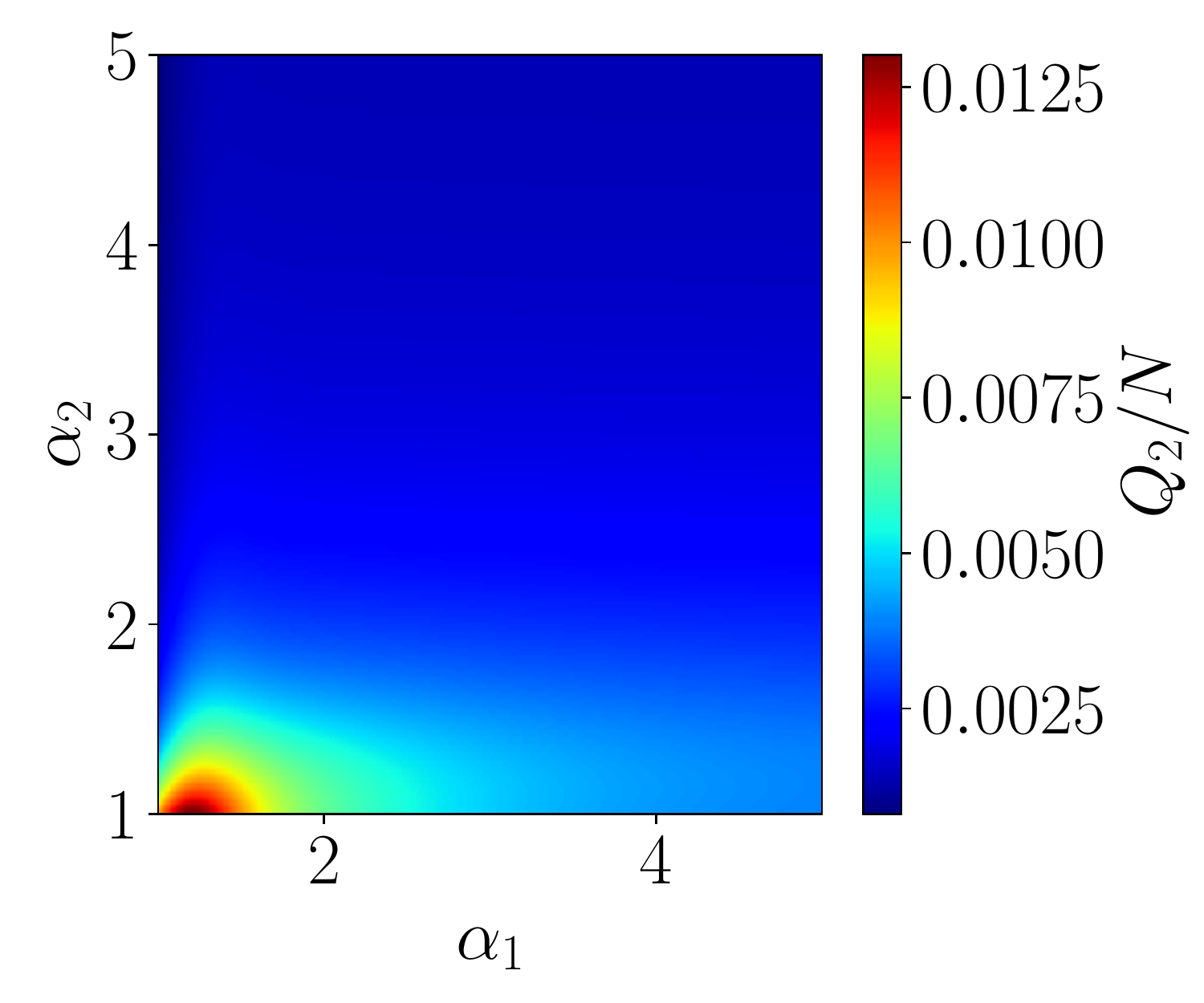}
			\caption{Heat exchanged with the cold reservoir plotted as a function of $\alpha_{1}$ and $\alpha_{2}$, for cycle parameters: $h_1 = 2$, $h_2 = -1+2^{1-\alpha_{1}}$, $T_1 = 0.1$, $T_2 \simeq 0.099$, and for system size $N = 200$.}
			\label{fig: Q2_a1_a2_hc}
		\end{figure}

		Analogously, in the refrigerator operation mode, the heat extracted from the cold bath $Q_2$ shows the same qualitative behavior independently of the values of $\alpha_{1}$ and $\alpha_{2}$. Figure \ref{fig: Q2_a12_h} shows $Q_2$ in the same parameter region as in Section \ref{sec: Refrigerator operation mode} of the main text, as a function of $h_2/h_1$ in the three extremal cases we considered also for the heat-engine. In particular, $Q_2$ is maximal when $h_2$ corresponds to one of the quantum critical points $h_2 = 1$ and $h_2 = -1+2^{1-\alpha_{1}}$, which are represented by the vertical dashed lines in the plots. However, we notice that the presence of long-range pairing amplitude $\alpha_{2}<\infty$ is necessary in order to have a clear long-range advantage for the cooling capability. Moreover, as shown in Fig. \ref{fig: Q2_a1_a2_hc}, the cooling capability $Q_2$ at the quantum critical points $h_2 = -1+2^{1-\alpha_1}$ is enhanced when both $\alpha_1$ and $\alpha_2$ are small.

		The analysis carried out in this Appendix justifies our choice to restrict the study presented in the main text to the most interesting case corresponding to equally long-range hopping and pairing amplitudes $\alpha_{1} = \alpha_{2} = \alpha$. 
	\section{Advantage-disadvantage transition}\label{app: Advantage-disadvantage transition}
	In this Appendix, we provide a detailed analysis of the advantage-disadvantage transition in the advantage parameter~\eqref{eq: advantage parameter E}, as the temperatures ratio $T_2/T_1$ departs from the optimal regime. Figure \ref{fig: DW_Scaling} shows $\Delta W$ as a function of $\beta_1/\beta_2$, for different values of the system size $N$. We notice that both the maximum advantage, $\max[\Delta W]$, and the maximum disadvantage $-\min[\Delta W]$, increase as $N$ grows. However, a finite advantage region with $\Delta W>0$ is always present at sufficiently small $\beta_1/\beta_2$, even in the thermodynamic limit. More precisely the finite size scaling of $\Delta W$ is of the form 
	\begin{align}
		\Delta W \approx \delta w N  
	\end{align}
	Consequently the transition point between long-range advantage and disadvantage is identified by the condition $\delta w = 0$, which happens at $\beta_1/\beta_2 = (\beta_1/\beta_2)^*$, independently from the system size, see the vertical black dashed line in Fig. \ref{fig: DW_Scaling}. 
	
	A good estimate of the transition point, $(\beta_1/\beta_2)^* = (T_2/T_1)^*$, can be obtained by considering the leading finite temperature corrections to Eq.~\eqref{eq: W0}, namely 
	\begin{align}
		W\simeq W_0 + \mathcal{W}(T_1,T_2).
	\end{align}
	Then in the limit $T_2\ll\bar{\omega}\ll T_1$, $\mathcal{W}(T_1,T_2)$ can be separated into two contributions 
	\begin{align}
		\mathcal{W}(T_1,T_2) = \mathcal{W}_1+ \mathcal{W}_2,
	\end{align}
	with
	\begin{align}
		\mathcal{W}_1 &= -\frac{\beta_1}{2}\sum_{k>0}\omega_{1,k}(\omega_{1,k}-\omega_{2,k})\\
		\mathcal{W}_2 &= -2\sum_{k>0}(\omega_{1,k}-\omega_{2,k})e^{-\beta_2\omega_{2,k}}.
	\end{align}
	We notice that $\mathcal{W}_1\propto\beta_1 = 1/T_1$, independently of the values of $h_1$ and $h_2$. On the other hand for $\mathcal{W}_2$, as $T_2\to 0$ ($\beta_2\to\infty$),if $h_2\neq h_{c}$, we have the exponential decay
	\begin{align}
		\mathcal{W}_2\approx e^{-\beta_2\min_k\omega_{2,k}}.
	\end{align}
	On the other hand, if the value of $h_2$ corresponds to one of the quantum critical points $h_2 = h_c$ we find a power law decay
	\begin{align}
		\mathcal{W}_2\approx T_2^{1/z},
	\end{align}
	where $z$ is the dynamical critical exponent~\eqref{eq: dynamical exponent 1},\eqref{eq: dynamical exponent 2}. It follows that the finite $T_2$ corrections are more important when $h_2 = -1+2^{1-\alpha_1}$, since in this case the dynamical critical exponent is $z = 1$. This corresponds to the advantage-disadvantage transition point with the smallest cold temperature $T_2 = T_2^*$ (see the white dashed line in Fig. \ref{fig: DW Deta}). In fact, the same finite temperature behavior is found also for the advantage parameter~\eqref{eq: advantage parameter E}, since in the $T_2\ll\bar{\omega}\ll T_1$ limit, we have
	\begin{align}
		\Delta W = \Delta W_0 + \Delta\mathcal{W}_1+\Delta\mathcal{W}_2,
	\end{align}
	where 
	\begin{align}
		\Delta\mathcal{W}_1\simeq -A_1/T_1,
	\end{align}
	and $\Delta\mathcal{W}_2$ has an exponential decay for $h_2 \neq h_c$, while at quantum criticality it has the power law decay
	\begin{align}
		\Delta\mathcal{W}_2\simeq -A_2T_2^{1/z}.
	\end{align}
	In particular, in order to determine the smallest transition temperature $T_2^*$, we take $h_2 = -1+2^{1-\alpha_1}$. Accordingly, we can write the advantage parameter as
	\begin{align}
		\Delta W \simeq \Delta W_0-A_1/T_1-A_2T_2.
	\end{align}
	Finally, imposing the transition condition $\Delta W = 0$ we obtain the transition temperature
	\begin{align}
		\frac{T_2^{*}}{T_1} \simeq \frac{W_0}{A_2} +\frac{A_1}{A_2}\frac{1}{T_1}\simeq \frac{W_0}{A_2}.
	\end{align}
	\section{Irreversible entropy production of the finite time cycle}\label{app: Entropy production}
	In this Appendix, we comment on the different sources of the cycle irreversibility by explicitly computing the cycle entropy production defined as $\Sigma = \beta_1Q_1+\beta_2Q_2$ in the finite-time case. The second law of thermodynamics, in the form of Clausius inequality, constrains this quantity to be non-positive $\Sigma\leq 0$, with the equality holding only if the cycle is perfectly reversible. Then, we can see $\Sigma$ as an indicator of the cycle irreversibility, which also tells us how much the performances of the device are close to the optimal one, represented by the Carnot bound. Interestingly, in our case the entropy production can be written as the sum of two contributions $\Sigma = \Sigma_{\infty}+\Sigma_{\delta}$. The first term, given by
	\begin{align}
		\Sigma_{\infty} = \sum_{k>0}(\beta_1\omega_{1,k}-\beta_2\omega_{2,k})\left[f_{2,k}-f_{1,k}\right],
	\end{align}
	is present also in the infinitely slow cycle and it is somehow unavoidable since it is due to the fact that the two thermalization strokes of the cycle are intrinsically irreversible. Thus $\Sigma_{\infty} = 0$ only exactly at the Carnot point, where however all the energies exchanges are null $Q_1 = Q_2 = W = 0$. On the other hand, the second contribution, reading
	\begin{align}
		\Sigma_{\delta} = \sum_{k>0}\left[\beta_1\omega_{1,k}f_{2,k}-\beta_2\omega_{2,k}f_{1,k}\right](1-P_k),\label{eq: nonadiabatic entropy production}
	\end{align}
	is present only when the unitary strokes are performed at a finite velocity. We notice that each term in the sum of Eq.~\eqref{eq: nonadiabatic entropy production} is proportional to the nonadiabatic transition probability $1-P_k$, then showing explicitly that they provide an additional source of irreversibility, resulting in worse efficiency performances in finite time cycles. In fact, comparing ~\eqref{eq: nonadiabatic entropy production} and ~\eqref{eq: W_infty-W optimal}, we notice that in the  $T_2\ll\bar{\omega}\ll T_1$ limit 
	\begin{align}
		W_\infty-W\simeq \beta_1\Sigma_\delta,
	\end{align}
	showing the relation between irreversible entropy production and nonadiabatic energy losses.
\end{document}